\documentclass[journal, a4paper]{IEEEtran}

\usepackage{amsmath,amssymb,mathrsfs,amsbsy}
\usepackage{cite}
\usepackage{subfigure}
\usepackage{graphicx,color}
\usepackage{hyperref}
\usepackage{subfigure}
\usepackage{balance}
\usepackage{tikz,siunitx}
\usepackage{url}
\usepackage{makecell}
\usepackage{algorithm,algorithmic}
\usepackage{multirow}

\newtheorem{lem}{Lemma}

\newtheorem{thm}{Theorem}

\DeclareMathOperator*{\argmin}{arg\,min}
\newcommand{\matc}[1]{\mbox{\boldmath $\mathcal{#1}$}}
\newcommand{\figref}[1]{Fig. \ref{#1}}

\definecolor{sblue}{RGB}{0,51,180}
\definecolor{sred}{RGB}{200,51,100}
\definecolor{seaBlue}{RGB}{0,105,148}

\ifCLASSINFOpdf
\else
\fi

\begin{document}

\title{Symmetric Rank-$1$ Regularization for Iterative Inversion of Ill-Conditioned MIMO Channels}

\author{Jinfei Wang, Yi Ma, and Rahim Tafazolli
\thanks{Jinfei Wang, Yi Ma, and Rahim Tafazolli are with the 6GIC, Institute for Communication Systems, University of Surrey, Guildford, United Kingdom, GU2 7XH, e-mails: (jinfei.wang, y.ma, r.tafazolli)@surrey.ac.uk. }
\thanks{This work has been partially presented in ICC'2023, Rome \cite{10278625}.}}
\markboth{IEEE Transactions on Wireless Communications (Draft)}%
{}

\maketitle

\begin{abstract}
While iterative matrix inversion methods excel in compatibility of parallel and distributed computing when managing large matrices, their limitations are also evident in multiple-input multiple-output (MIMO) fading channels.
These methods encounter challenges related to slow convergence and diminished accuracy, especially in ill-conditioned scenarios, hindering their application in future MIMO networks such as extra-large aperture array.
To address these challenges, this paper proposes a novel matrix regularization method termed symmetric rank-$1$ regularization (SR-$1$R).
The proposed method functions by augmenting the channel matrix with a symmetric rank-$1$ matrix, with the primary goal of minimizing the condition number of the resultant regularized matrix.
This significantly improves the matrix condition, enabling fast and accurate iterative inversion of the regularized matrix.
Then, the inverse of the original channel matrix is obtained by applying the Sherman-Morrison transform on the outcome of iterative inversions.
Our eigenvalue analysis unveils the best channel condition that can be achieved by an optimized SR-$1$R matrix.
Moreover, a power iteration-assisted (PIA) approach is proposed to find the optimum SR\nobreakdash-$1$R matrix without need of eigenvalue decomposition.
The proposed approach exhibits logarithmic algorithm-depth in parallel computing for MIMO precoding. 
Finally, computer simulations demonstrate that SR-$1$R has the potential to reduce the required iteration by up to $35\%$ while achieving the performance of regularized zero-forcing.
\end{abstract}


\section{Introduction}\label{secI}


\IEEEPARstart{I}{terative} matrix inversion methods play a pivotal role in scenarios where direct methods become impractical or incurs prohibitive computational latency.
This significance becomes especially pronounced when handling large matrices, as iterative techniques offer superior compatibility to parallel and distributed computing as well as enhanced numerical stability \cite{Schulz1933,Hotelling1943,GowerThesis}.
In addition, as future multiple-input multiple-output (MIMO) hardware increasingly incorporates high parallel computing capabilities (e.g., quantum computing \cite{Zhang2016,Norimoto2023}), such compatibility will be essential for enabling scalable and efficient signal processing algorithms.

Despite their promise, current iterative matrix inversion methods face difficulties with slow convergence and reduced accuracy as the wireless channel becomes ill-conditioned in future multi-user MIMO systems.
Driven by the ever growing demand for spectral efficiency, user terminals are equipped with more antennas; this is pushing multi-user MIMO back to (near)~symmetric architecture where the channel is extremely ill-conditioned \cite{Wang2022c}.
With the increase of user antennas, increasing the service antennas to maintain the service-to-user antenna ratio as in massive MIMO systems becomes hardware costing and is challenging to device manufactory for the calibration of antenna array \cite{6375940}.
Moreover, when the operating frequency is fixed, adding more antennas often means enlarging the aperture of antenna-array. 
This could drag user terminals into the near field of the transmitter and transform conventional MIMO systems into an extra-large aperture array (ELAA) MIMO (see \cite{BJORNSON20193,Wu2023a} for the concept).
The ELAA-MIMO features significant channel spatial non-stationarity on the network side and exacerbates the ill-conditioning of wireless channel beyond what is typical in conventional MIMO systems \cite{10295381,10653741,Wang2022b,Wu2023}.


Basically, the channel condition can be improved by means of matrix regularization, pre-conditioning or a combination of them.
The regularized zero-forcing (RZF) stands as a widely used regularization method, yet our research reveals its subpar performance in iterative channel inversion (see Section \ref{secV}).
In contrast, pre-conditioning involves multiplying a specialized matrix with the original channel matrix, often formed from a sub-matrix inverse \cite{8417575,Lee2020,7399337,Saad2003}.
This approach enhances matrix conditions, leading to improved iterative inversion in numerous classical methods like Jacobi \cite{8417575}, Gauss-Seidel (GS) \cite{Lee2020}, and symmetric successive over-relaxation (SSOR) \cite{7399337,Saad2003}.
However, simulation results elaborated in Section \ref{secV} unveils their inability to deliver satisfactory performance in ill-conditioned {MIMO} channels.

\begin{figure}[t]
\centering
\includegraphics[scale=0.3]{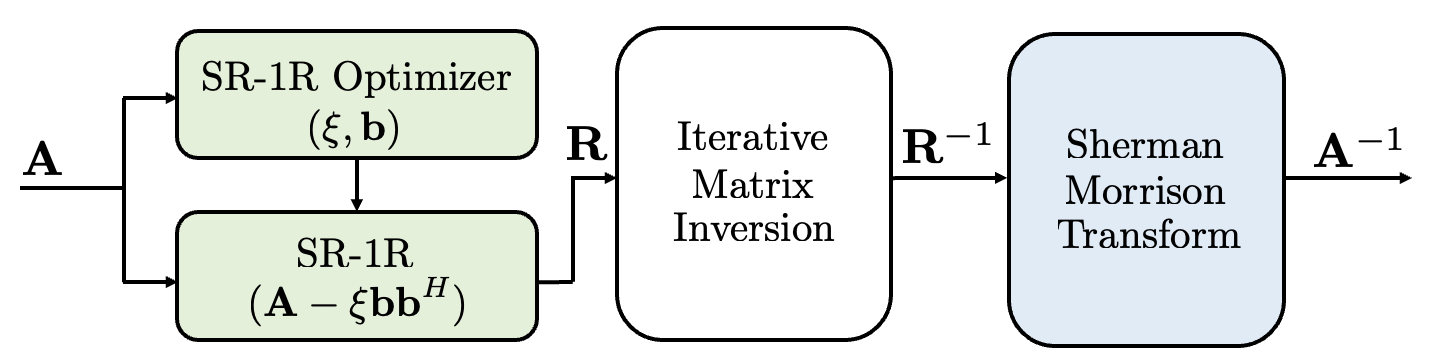}
\caption{Block diagram of the proposed symmetric rank-$1$ regularization for iterative matrix inversion.}
\label{SR1R}
\vspace{-2em}
\end{figure}

This paper introduces a novel channel matrix regularization approach, termed symmetric rank-$1$ regularization (SR-$1$R). 
As per the block diagram illustrated in \figref{SR1R}, we consider a positive definite matrix $\mathbf{A}\in\mathbb{C}^{N\times N}$, where $N$ represents the matrix size.
Rather than directly computing the inverse matrix $\mathbf{A}^{-1}$, the aim of SR-$1$R is to firstly find the inverse of a rank-$1$ regularized matrix $\mathbf{R}\in\mathbb{C}^{N\times N}$ defined as:
\begin{equation}\label{eqn01}
\mathbf{R}\triangleq\mathbf{A}-\xi\mathbf{b}\mathbf{b}^H,
\end{equation}
where $\xi\in\mathbb{R}$ is a scaling factor, $\mathbf{b}\in\mathbb{C}^{N\times1}$ is a unitary vector (i.e., $\|\mathbf{b}\|=1$), $[\cdot]^H$ and $\|\cdot\|$ denotes the Hermitian transpose and the Euclidean norm, respectively.
Next, Sherman-Morrison transform (see \cite{Petersen2012}) is applied on $\mathbf{R}^{-1}$, resulting in:
\begin{equation}\label{eqn02}
\mathbf{A}^{-1}=\mathbf{R}^{-1}-\frac{\xi\mathbf{R}^{-1}\mathbf{b}\mathbf{b}^H\mathbf{R}^{-1}}{1+\xi\mathbf{b}^H\mathbf{R}^{-1}\mathbf{b}}.
\end{equation}
Our research aims to ascertain the existence of $(\xi, \mathbf{b})$ values that could improve the condition of $\mathbf{R}$ compared to $\mathbf{A}$. If such values exist, our goal is to optimize $(\xi, \mathbf{b})$ to maximize the channel condition.

An intuitive explanation of the SR-$1$R is to reduce the largest eigenvalue below the second largest one while simultaneously increasing the modulus of the smallest eigenvalue above the second smallest one.	
This design is of interest because the largest and smallest eigenvalues can heavily influence the channel ill-conditioning. 
For instance, the largest eigenvalue dominates the ill-conditioning when the channel is highly correlated due to line-of-sight (LoS) links, while the smallest eigenvalue can make the channel condition unstable in (near) symmetric MIMO.	
Notably, the SR-$1$R design is primarily targeted toward ill-conditioned channels. 
In typical massive MIMO systems, where the channel condition is generally favorable, the largest and smallest eigenvalues are close to their neighbors.
This limits the gain brought by the SR-$1$R.
Therefore, such scenarios are outside the primary scope of this design.

Alongside the novel SR-$1$R concept, major contributions of our research include:

\subsubsection{Feasibility and optimality study of SR-$1$R}
Our study, through eigenvalue analysis, reveals the presence of an optimal $(\xi, \mathbf{b})$ combination that maximizes the matrix condition of $\mathbf{R}$.
Detailed results and justifications are referred to Section \ref{secIII}.

\subsubsection{Practical approaches for optimization}
While theoretically, optimal determination of $(\xi, \mathbf{b})$ involves eigenvalue decomposition (EVD) or singular value decomposition (SVD) of $\mathbf{A}$, these methods conflict with the purpose of employing iterative matrix inversion, particularly in wireless MIMO scenarios where direct matrix inversion is unfeasible.

To tackle this challenge, our research reveals that $(\xi, \mathbf{b})$ can be optimally established solely by discerning the largest and smallest eigenvalues of $\mathbf{A}$ and their corresponding eigenvectors. 
Subsequently, we employ a power iteration technique to compute the largest eigenvalue of $\mathbf{A}$ and a shifted power iteration method to compute the smallest eigenvalue, concurrently obtaining their respective eigenvectors.
It is shown that sufficiently accurate estimates of eigenvalues and eigenvectors can be obtained through just one iteration.

\subsubsection{Simulations and Performance Evaluation}
The SR-$1$R approach has undergone extensive simulations and performance evaluation across a range of MIMO fading channel types, encompassing identical and independently distributed (i.i.d.) symmetric MIMO Rayleigh channels, i.i.d symmetric MIMO Rician channels, LoS/non-LoS mixed ELAA-MIMO channels, and LoS-dominated ELAA-MIMO channels \cite{10295381}. These channels exhibit diverse and contrasting channel conditions. Our simulation results consistently demonstrate that the SR-$1$R approach substantially enhances the channel conditions in all of these channel models.
Thanks to this remarkable advantage, the SR-$1$R-based iterative channel inversion exhibits a significant performance enhancement in MIMO linear precoding, reducing the required iterations by more than $4\%$ when achieving RZF performance.
In the case of the i.i.d. MIMO Rician channel, the reduction in the number of iterations even reaches $35\%$.
Moreover, the SR-$1$R demonstrates outstanding compatibility with parallel computing, showing only minor changes in computation latency when the MIMO size increases dramatically.


\section{System Model, Prior Art, and Problem Analysis}\label{secII}
The SR-$1$R approach exhibits broad applicability across various scenarios necessitating iterative matrix inversion. 
However, our paper specifically concentrates on its implementation within wireless MIMO channel inversion, particularly tailored to MIMO linear precoding technology. 
This focus is motivated by the prevalent significance of MIMO linear precoding in advanced wireless communication systems, where channel inversion holds critical role for optimizing signal transmission efficiency and reducing interference, making it an ideal area for SR-$1$R exploration.

\subsection{MIMO Linear Precoding}\label{sec21}
Consider MIMO downlink communications, where an access point (i.e., the transmitter) equipped with $M$ transmit-antennas sends $N$ independent data streams to $N$ $(\leq M)$ user antennas, respectively. 
Denote $\mathbf{s}\in\mathbb{C}^{N\times 1}$ as an information-bearing symbol block in the spatial domain. 
To maintain the orthogonality between data streams, an {interference-rejection linear precoder}, denoted as $\mathbf{W}\in\mathbb{C}^{M\times N}$, is employed at the transmitter, which yields the transmit symbol block:  
\begin{equation}\label{eq03}
\mathbf{x}=\mathbf{W}\mathbf{s}.
\end{equation}
The symbol block received through the MIMO channel $(\mathbf{H}\in\mathbb{C}^{N\times M})$ is represented by:
\begin{equation}\label{eq04}
\mathbf{y}=\sqrt{\mathcal{E}}\mathbf{H}\mathbf{x}+\mathbf{v},
\end{equation}
where $\mathbf{y}\in\mathbb{C}^{N\times1}$ stands for the received block, 
$\mathcal{E}$ for the transmit power, 
$\mathbf{v}\sim\mathcal{CN}(\mathbf{0},\sigma_v^2\mathbf{I}_N)$ for the additive white Gaussian noise (AWGN),
and $\mathbf{I}_N$ for the identity matrix with the dimension of $N$.

A commonly employed {interference-rejection linear precoder} is known as the zero-forcing (ZF) precoder \cite{1468466}, i.e.,
\begin{equation}\label{eq05}
\mathbf{W}\triangleq\mathbf{H}^H\mathbf{A}^{-1},~ \mathbf{A}\triangleq\mathbf{H}\mathbf{H}^H\in\mathbb{C}^{N\times N},
\end{equation}
where the Gram matrix $\mathbf{A}$ is positive definite. 
Applying \eqref{eq05} into \eqref{eq04} yields
\begin{equation}\label{eq06}
\mathbf{y}=\sqrt{\mathcal{E}}\mathbf{s}+\mathbf{v}.
\end{equation}
With the ZF precoder, data streams do not interfere with each other and can be decoded separately. 
On the other hand, the computation of matrix inverse $\mathbf{A}^{-1}$ costs cubic complexity and does not support well the parallel computing technology \cite{Yang2015}.
This presents a formidable challenge in the design of scalable large-MIMO systems.

In the realm of MIMO, it is important to note that RZF is regarded as a prominent {interference-rejection linear precoder} approach. 
However, in this paper, we will explore RZF as a regularization method in Sec. \ref{sec2b}.

Notably, the channel ill-conditioning increases the required SNR of interference-rejection linear precoding.
This issue can be effectively mitigated by employing non-linear precoding \cite{Wang2022c,Li2023Group}.
The non-linear precoding also requires fast and accurate channel inversion, further motivating our study.
In this paper, we focus our study on iterative linear precoding.	

\subsection{Prior Art Analysis}\label{sec2b}
Our analysis of prior art identifies two main categories of iterative solutions to the inversion problem: 
{\em a)} iterative LSI, and {\em b)} iterative matrix inversion. 
This subsection briefly explores both categories, laying the groundwork for advocating the necessity of employing iterative matrix inversion.

\subsubsection{Iterative LSI}
In numerous use cases, the matrix inverse problem can be circumvented by transforming it into a LSI problem.
Specifically for the linear ZF precoding problem \eqref{eq05}, rather than directly computing $\mathbf{A}^{-1}$, we can find a vector \mbox{$\overline{\mathbf{x}}\in\mathbb{C}^{N\times 1}$} that is related to $\mathbf{x}$ through the equation: $\mathbf{x}=\mathbf{H}^H\overline{\mathbf{x}}$.
In line with \eqref{eq05}, the goal is to iteratively solve the following equation (e.g., \cite{Wang2023Hanzo,Goldfarb1972,Richardson1911}): 
\begin{equation}\label{eq07}
\mathbf{A}\overline{\mathbf{x}}=\mathbf{s}.
\end{equation}
An analogous approach can also be found in the MIMO uplink detection (e.g., \cite{Wang2022Gower}).


Iterative LSI approaches reduce the computational complexity of matrix–matrix multiplication to matrix-vector multiplications.
Moreover, good initialization can substantially enhance convergence when the number of iterations is limited (e.g., around $10$). 
These properties make iterative LSI approaches particularly attractive under well-conditioned channel scenarios, such as in massive MIMO systems, as demonstrated in recent studies \cite{Wang2022Gower,Li2022a}.

When the channel is ill-conditioned, however, the required number of iterations increases, and the convergence speed is more critical. Iterative LSI methods generally exhibit linear convergence speed, whereas iterative matrix inversion algorithms can achieve exponential convergence []. 
Therefore, in our following discussions, we will focus our study on iterative matrix inversion.

\subsubsection{Iterative matrix inversion}
Various techniques are available for the iterative matrix inversion. Common methods include the Neumann series method \cite{8064675} and the Schulz method \cite{Schulz1933}. 
The Schulz method is also known as the Hotelling-Bodewig algorithm \cite{Hotelling1943} or the Newton iteration \cite{Wang2022Gower} in the literature.
Additionally, the sketch-and-projection approach has been recently adopted for matrix inversion \cite{GowerThesis}. 
Our state-of-the-art analysis concludes that the Neumann series method typically requires more iterations compared to the Schulz method \cite{Tang2016}. Furthermore, the performance of the sketch-and-projection method is influenced by the randomness of the sketching process, making it less stable than the Schulz method.
Therefore, in this paper, we opt for the Schulz method to evaluate the performance of iterative matrix inversion concerning the condition numbers $\kappa(\mathbf{A})$ or $\kappa(\mathbf{R})$.

Specifically, considering $\mathbf{X}_i$ as the output of the $i^{th}$ iteration of iterative matrix inversion, the convergence criterion for an iterative algorithm is represented by the Frobenius norm converging as:
\begin{equation}\label{eq08}
\lim_{i\to\infty}\|\mathbf{I}-\mathbf{A}\mathbf{X}_{i}\|_\mathrm{F}=0,
\end{equation}
i.e., $\mathbf{X}_\infty=\mathbf{A}^{-1}$.
To achieve this goal, the Schulz method suggests the following iterative process:
\begin{equation}\label{eq09}
\mathbf{X}_0=\omega\mathbf{A}^H;~\mathbf{X}_{i}=2\mathbf{X}_{i-1} - \mathbf{X}_{i-1}\mathbf{A}\mathbf{X}_{i-1},~i>0,
\end{equation}
where $\omega$ is a scaling factor ensuring the condition
\begin{equation}\label{eq10}
\rho(\mathbf{I}-\omega\mathbf{A}\mathbf{A}^H)<1,
\end{equation}
and $\rho(\cdot)$ is the spectrum radius of a matrix.


\subsection{Research Problem and Analysis}
For the iterative matrix inversion, the number of iterations should be minimized while ensuring satisfactory performance, as this is crucial for maximizing computational efficiency and accuracy.
The number of iterations is strongly dependent on the condition number $\kappa(\mathbf{A})$.
For instance, in the case of a unitary channel matrix with $\kappa(\mathbf{A})=1$, 
setting $\omega=1$ in \eqref{eq09} immediately yields $\|\mathbf{I}-\mathbf{A}\mathbf{X}_{0}\|_\mathrm{F}=0$.

As far as MIMO fading channels are concerned, it is empirically evident that the number of iterations grows rapidly as the condition number $\kappa(\mathbf{A})$ increases.

This challenge is particularly pronounced in the case of ELAA-MIMO channels, which often experience poorer channel conditions compared to traditional MIMO channels due to their network-side spatial channel inconsistency. Hence, there is a pressing need to find an efficient approach to enhance the condition number of ELAA-MIMO channels

The current approaches for improving the condition of the channel matrix mainly involve matrix regularization and preconditioning techniques. A widely recognized matrix regularization method is known as the RZF, which performs the iterative inversion on the regularized matrix
\begin{equation}\label{eqn10}
\mathbf{R}_\textsc{rzf}=(\mathbf{A}+\rm{snr}^{-1}\mathbf{I}),
\end{equation}
where $\rm{snr}$ denotes the signal-to-noise ratio (SNR).
Our simulation results, presented in Section \ref{secV}, demonstrate that the RZF method generally does not provide satisfactory performance when applied to invert ELAA-MIMO channels. 
Notably, RZF exhibits particularly poor performance at high SNRs.
This is easily understandable because, under high SNRs, $\mathbf{R}_\textsc{rzf}$ is almost identical to $\mathbf{A}$.

For the pre-conditioning method, the channel matrix $\mathbf{A}$ is multiplied with a pre-conditioning matrix $\mathbf{P}^{-1}$, which yields $\mathbf{R}_\textsc{pre}=\mathbf{P}^{-1}\mathbf{A}$.
Iterative inversion is performed on $\mathbf{R}_\textsc{pre}$, which aims to generate $\mathbf{R}_\textsc{pre}^{-1}=\mathbf{A}^{-1}\mathbf{P}$.
After the iterative inversion, $\mathbf{A}^{-1}$ is computed through $\mathbf{A}^{-1}=\mathbf{R}_\textsc{pre}^{-1}\mathbf{P}^{-1}$.

Prior art pre-conditioning methods mainly include the Jacobi pre-conditioning \cite{8417575}, the GS pre-conditioning \cite{Lee2020,Wang2023Hanzo}, and the SSOR pre-conditioning \cite{Saad2003}.
Their major difference lies in the formation of the pre-conditioning matrix $\mathbf{P}$, i.e.,
\begin{IEEEeqnarray}{rl}
\rm{Jacobi}:\quad& \mathbf{P}=\mathbf{D};\label{eq14New}\\
\rm{GS}:\quad& \mathbf{P}=\mathbf{D}+\mathbf{L};\label{eq15New}\\
\rm{SSOR}:\quad& \mathbf{P}=(\mathbf{D}+\mathbf{L})\mathbf{D}^{-1}(\mathbf{D}+\mathbf{L})^H,\label{eq16New}
\end{IEEEeqnarray}
where $\mathbf{D}$ is a diagonal matrix formed by the diagonal of $\mathbf{A}$, and $\mathbf{L}$ is the strict lower triangular part of $\mathbf{A}$. 
It is perhaps worth noting that inverting a lower-triangular matrix entails a computational complexity that scales quadratically, which keeps both the GS and SSOR approaches within the realm of low complexity methods.

Despite extensive prior research documented in the literature, pre-conditioning approaches have only yielded marginal performance enhancements when dealing with ill-conditioned ELAA-MIMO channels (as detailed in Section \ref{secV}). 
To further enhance both the performance and convergence in iterative matrix inversion, we introduce the SR-$1$R method.

\section{SR-$1$R: Feasibility and Optimality Study}\label{secIII}
The concept of SR-$1$R has already been introduced in Section \ref{secI}, as outlined in Eqn. \eqref{eqn01}-\eqref{eqn02}.
In this section, our objective is to address the question of whether there exists $(\xi, \mathbf{b})$ that can lead to: {\em a)} $\kappa(\mathbf{R})<\kappa(\mathbf{A})$; {\em b)} minimization of $\kappa(\mathbf{R})$.
This question motivates the feasibility and optimality study, relying on the assumption of availability of EVD (or SVD as appropriate).
This assumption is reasonable here as it is specifically for feasibility and optimality analysis.

As a quick overview of our finding, upon defining eigenvalues of the positive-definite matrix $\mathbf{A}$ as $(\lambda_0, \lambda_1, ..., \lambda_{N-1})$, arranged in descent order, our mathematical analysis reveals the existence of $(\xi, \mathbf{b})$ yielding: 
\begin{equation}\label{eq17}
\min\big(\kappa(\mathbf{R})\big)=\frac{\lambda_1}{\lambda_{N-2}}.
\end{equation}
This condition number is smaller than the condition number of $\mathbf{A}$
\begin{equation}\label{eq18}
\kappa(\mathbf{A})=\frac{\lambda_0}{\lambda_{N-1}},
\end{equation}
signifying a notable improvement.
Further elaboration and detailed proof of this finding are provided in the subsequent subsections.

\subsection{Feasibility Study}\label{secIIIa}
Similar to $\mathbf{A}$, Eqn. \eqref{eqn01} shows that $\mathbf{R}$ is also a symmetric matrix. 
However, $\mathbf{R}$ may not exhibit positive-definite properties, introducing a degree of complexity into the mathematical analysis of its condition number.

To facilitate our analysis, we express the regularized matrix $\mathbf{R}$ as the following form
\begin{equation}\label{eq19}
\mathbf{R}=\mathbf{U}(\mathbf{\Lambda}-\xi\mathbf{p}\mathbf{p}^H)\mathbf{U}^H,
\end{equation}
where $\mathbf{\Lambda}\triangleq\rm{diag}\{\lambda_0, \lambda_1, ..., \lambda_{N-1}\}$, $\mathbf{U}\in\mathbb{C}^{N\times N}$ is the unitary matrix coming from the EVD: $\mathbf{A}=\mathbf{U\Lambda U}^H$, and $\mathbf{p}\triangleq\mathbf{U}^H\mathbf{b}$.

Consider the matrix $\mathbf{\Theta}$ defined as 
\begin{equation}\label{eq20}
\mathbf{\Theta}\triangleq\mathbf{\Lambda}-\xi\mathbf{p}\mathbf{p}^H.
\end{equation}
It is worthwhile to note that $\mathbf{\Theta}$ and $\mathbf{R}$ possess identical eigenvalues, and consequently have $\kappa(\mathbf{\Theta})=\kappa(\mathbf{R})$. 
Hence, our investigation can be directed towards the analysis of $\mathbf{\Theta}$. 

In the mathematical literature, \eqref{eq20} is well known as the rank-$1$ modification of symmetric eigenproblem \cite{Gu1994}. 
One of the significant findings from previous research can be  summarized as:
\begin{lem}[from \cite{Gu1994}]\label{lem01}
Define the eigenvalues of the rank-$1$ modified symmetric matrix $\mathbf{\Theta}$ as $(\theta_0, \theta_1, ..., \theta_{N-1})$.
When $\xi<0$, $\mathbf{\Theta}$ is a positive-definite matrix whose eigenvalues satisfies the following inequalities 
\begin{IEEEeqnarray}{ll}\label{eq21}
&\lambda_{0}\leq\theta_0\leq\lambda_{0}-\xi,\nonumber\\
&\lambda_{n}\leq\theta_{n}\leq\lambda_{n-1},~_{n=1,\cdots,N-1}.
\end{IEEEeqnarray}
When $\xi>0$, the following inequalities are satisfied
\begin{IEEEeqnarray}{ll}\label{eq22}
&\lambda_{N-1}-\xi\leq\theta_{N-1}\leq\lambda_{N-1},\nonumber\\
&\lambda_{n+1}\leq\theta_{n}\leq\lambda_{n},~_{n=0,\cdots,N-2}.
\end{IEEEeqnarray}
\end{lem}

According to {\em Lemma \ref{lem01}}, when $\xi<0$, the matrix $\mathbf{\Theta}$ is positive-definite, and its eigenvalues coincide with its singular values. The condition number of $\mathbf{\Theta}$ can be determined using \eqref{eq21}, expressed as
\begin{equation}\label{eq23}
\kappa(\mathbf{\Theta})=\frac{\theta_0}{\theta_{N-1}}\in\left[\frac{\lambda_0}{\lambda_{N-2}}, \frac{\lambda_0-\xi}{\lambda_{N-1}}\right].
\end{equation}
However, in this case, it cannot be guaranteed that the inequality $\kappa(\mathbf{\Theta}) < \kappa(\mathbf{A})$ holds. Consequently, having $\xi<0$ is not conducive for the SR-$1$R approach.

For the case of $\xi>0$, $\mathbf{\Theta}$ is not necessarily a positive-definite matrix, and thus its eigenvalues do not coincide with its singular values. 
As per \eqref{eq22}, the only inconsistency comes from $\theta_{N-1}$, which can be either positive or negative ($\xi\neq\lambda_{N-1}$). 
Therefore, the smallest singular value of $\mathbf{\Theta}$, denoted as $\overline{\theta}_{\min}$, is 
\begin{equation}\label{eq24}
\overline{\theta}_{\min}=\min(|\theta_{N-1}|, \theta_{N-2}),
\end{equation}
and the largest singular value of $\mathbf{\Theta}$, denoted as $\overline{\theta}_{\max}$, is 
\begin{equation}\label{eq25}
\overline{\theta}_{\max}=\max(|\theta_{N-1}|, \theta_{0}),
\end{equation}
where $|\cdot|$ stands for the absolute value of a number. 
Then, the condition number of $\mathbf{\Theta}$ can be expressed as
\begin{equation}\label{eq26}
\kappa(\mathbf{\Theta})=\frac{\overline{\theta}_{\max}}{\overline{\theta}_{\min}}
=\frac{\max(|\theta_{N-1}|, \theta_{0})}{\min(|\theta_{N-1}|, \theta_{N-2})}.
\end{equation}

To analyze \eqref{eq26} more thoroughly, we must rely on the following result:
\begin{thm}\label{thm01}
When $\mathbf{\Theta}$ is neither positive-definite nor positive semi-definite, 
there exists a $\xi(>0)$ such that the smallest eigenvalue of $\mathbf{\Theta}$ (i.e., $\theta_{N-1}$) satisfies 
\begin{equation}\label{eq27}
-\theta_{0}<\theta_{N-1}<-\theta_{N-2}.
\end{equation}
\end{thm}

\begin{IEEEproof}
We prove a sufficient condition of \textit{Theorem~\ref{thm01}}: $\theta_{N-1}$ monotonically decreases from $0$ to $-\infty$ as $\xi$ increases from $\xi^\bot$ to $\infty$, where $\xi^\bot$ is a positive scalar explained later in \eqref{eq05proof}.

To facilitate our study, define $f(\theta,\xi)$ as
\begin{equation}
f(\theta,\xi)\triangleq\det(\mathbf{\Theta}-\theta\mathbf{I})=\det(\mathbf{\Lambda}-\xi\mathbf{p}\mathbf{p}^H-\theta\mathbf{I}),
\end{equation}
where $\det(\cdot)$ stands for the determinant of a matrix.
$\theta_{N-1}$ is the smallest root of $f(\theta,\xi)=0$.

After some tedious mathematical work, the closed form expression of $f(\theta,\xi)$ is given by
\begin{IEEEeqnarray}{rl}\label{eq03proof}
f(\theta,\xi)&=(-\theta)^N\IEEEnonumber\\
+&\sum_{n=0}^{N-1}(-\theta)^{N-1-n}\Big(e_{n+1}(\mathbf{\Lambda})-\xi\sum_{k=0}^{N-1}e_{n}(\overline{\mathbf{\Lambda}}_k)|p_k|^2\Big),~~~~
\end{IEEEeqnarray}
where $e_{n+1}(\mathbf{\Lambda})$ stands for the $(n+1)^{th}$ order elementary symmetric function (ESF) formed by $\lambda_{0},\cdots,\lambda_{N-1}$ and $e_{n}(\overline{\mathbf{\Lambda}}_k)$ for the $n^{th}$ order ESF formed by $\lambda_{0},\cdots,\lambda_{k-1},\lambda_{k+1},\cdots,\lambda_{N-1}$ \cite{Macdonald1998}.
Since $\lambda_{n}>0$ $\forall n$, we have $e_{n+1}(\mathbf{\Lambda})>0$ and $e_{n}(\overline{\mathbf{\Lambda}}_k)>0$ $\forall n,k$.



The proof is divided into three steps.
In the first step, we prove
\begin{equation}\label{eq04proof}
\theta_{N-1}<0~\mathrm{for}~\xi>\xi^\bot,
\end{equation}
where $\xi^\bot$ is given by
\begin{equation}\label{eq05proof}
\xi^\bot=\frac{e_N(\mathbf{\Lambda})}{\sum_{k=0}^{N-1}e_{N-1}(\overline{\mathbf{\Lambda}}_k)|p_k|^2}.
\end{equation}
Substituting \eqref{eq05proof} into \eqref{eq03proof} when $\theta=0$ yields $f(0,\xi)=e_N(\mathbf{\Lambda})(1-\xi/\xi^\bot)$.
Thus, we have
\begin{equation}
f(0,\xi)<0~\mathrm{for}~\xi>\xi^\bot.
\end{equation}
When $\theta\to-\infty$, $f(\theta,\xi)$ is dominated by the term with the highest order:
\begin{equation}
\lim\limits_{\theta\to-\infty}f(\theta,\xi)=\lim\limits_{\theta\to-\infty}(-\theta)^N=\infty.
\end{equation}
Since $\theta_{N-1}$ is the only root within $(-\infty,\lambda_{N-1})$, it must fall into $(-\infty,0)$ for $\xi>\xi^\bot$ according to the intermediate value theorem.
\eqref{eq04proof} is proved.

In the second step, we prove that $\theta_{N-1}$ monotonically decreases with the increase of $\xi$ when $\xi>\xi^\bot$, i.e.,
\begin{equation}\label{eq07proof}
\theta_{N-1}^{(2)}<\theta_{N-1}^{(1)}<0~\mathrm{for}~\xi^{(2)}>\xi^{(1)}>\xi^\bot.
\end{equation}
This monotonicity can be proved by studying the following partial derivative:
\begin{equation}\label{eq08proof}
\frac{\partial f(\theta,\xi)}{\partial\xi}=\sum_{n=0}^{N-1}(-\theta)^{N-1-n}\Big(-\sum_{k=0}^{N-1}e_{n}(\overline{\mathbf{\Lambda}}_k)|p_k|^2\Big).
\end{equation}
When $\theta<0$, \eqref{eq08proof} is negative.
Since $\theta_{N-1}^{(1)}<0$, the following inequality holds
\begin{equation}\label{eq09proof}
f(\theta_{N-1}^{(1)},\xi^{(2)})<f(\theta_{N-1}^{(1)},\xi^{(1)})=0.
\end{equation}
\eqref{eq09proof} indicates $\theta_{N-1}^{(2)}\in(-\infty,\theta_{N-1}^{(1)})$ according to the intermediate value theorem, and proves \eqref{eq07proof}.

The first two steps prove that $\theta_{N-1}(<0)$ monotonically decreases with the increase of $\xi(>\xi^\bot)$.
In the final step, we prove that
\begin{equation}\label{apdx12}
\lim\limits_{\xi\to\infty}\theta_{N-1}=-\infty.
\end{equation}
Since $\|\mathbf{p}\|=1$, we have
\begin{equation}\label{apdx13}
    \sum_{n=0}^{N-1}\theta_n=\mathrm{Tr}(\mathbf{\Theta})=\mathrm{Tr}(\mathbf{\Lambda}-\xi\mathbf{p}\mathbf{p}^H)=\sum_{n=0}^{N-1}\lambda_{n}-\xi.
\end{equation}
The right hand side of \eqref{apdx13} tends to $-\infty$ as $\xi\to\infty$.
Recall \eqref{eq22} that $\theta_{n,}$ $_{n=0,\cdots,N-2}$ is bounded. 
Hence, we must have $\theta_{N-1}\to-\infty$ as $\xi\to\infty$ in the left hand side.
\eqref{apdx12} is proved.

Therefore, $\theta_{N-1}$ monotonically decreases from $0$ to $-\infty$ as $\xi$ increases from $\xi^\bot$ to $\infty$, yielding \textit{Theorem~\ref{thm01}}.
\end{IEEEproof}

The result \eqref{eq27} immediately leads to the inequality
\begin{equation}\label{eq28}
\theta_{0}>|\theta_{N-1}|>\theta_{N-2}.
\end{equation}
Applying \eqref{eq28} and \eqref{eq22} into \eqref{eq26} yields
\begin{equation}\label{eq29}
\kappa(\mathbf{R})=\kappa(\mathbf{\Theta}){=}\frac{\theta_0}{\theta_{N-2}}\leq\frac{\lambda_0}{\lambda_{N-1}}=\kappa(\mathbf{A}).
\end{equation}
This intermediate result ensures the feasibility of employing SR-$1$R to improve the matrix condition. 
Nevertheless, an in-depth study is needed to establish the optimality of SR-$1$R. 

\subsection{Optimality Study}\label{secIIIb}
In addition to the upper bound in \eqref{eq29}, as per \eqref{eq22}, it is evident that $\theta_0\geq \lambda_1$ and $\theta_{N-2}\leq\lambda_{N-2}$, and \eqref{eq26} yields also the following lower bound
\begin{equation}\label{eq30}
\kappa(\mathbf{\Theta})\geq\frac{\lambda_1}{\lambda_{N-2}}.
\end{equation}
This result indicates the best matrix condition that can be possibly achieved via the SR-$1$R approach. 

From a mathematical perspective, determining the optimal configuration of $(\xi, \mathbf{p})$ in a general form is extremely challenging. 
Nevertheless, we have successfully identified a special case of $(\xi, \mathbf{p})$ that is optimum, achieving the lower bound specified in \eqref{eq30}.

In this specific case, the vector $\mathbf{p}$ are configured as:
\begin{equation}\label{eq31}
\mathbf{p}=[\alpha, 0, ...,0, \beta]^T, \alpha,\beta\neq 0, \alpha^2+\beta^2=1.
\end{equation}
Applying \eqref{eq31} into \eqref{eq20} results in
\begin{equation}\label{eq32}
\mathbf{\Theta}=\left[
\begin{array}{ccccc}
\lambda_0-\xi\alpha^2&0&\cdots&0&-\xi\alpha\beta\\
0&\lambda_1& 0& \cdots & 0\\
\vdots& \vdots& \ddots& \vdots&\vdots\\
0&0&\cdots&\lambda_{N-2}&0\\
-\xi\alpha\beta& 0&\cdots & 0&\lambda_{N-1}-\xi\beta^2
\end{array}
\right]
\end{equation}
Immediately, we can identify $(\lambda_1, \lambda_2, ..., \lambda_{N-2})$ as eigenvalues of $\mathbf{\Theta}$.
The remaining two eigenvalues stem from a sub-matrix of $\mathbf{\Theta}$, constituted by:
\begin{equation}\label{eq33}
\overline{\mathbf{\Theta}}=\left[
\begin{array}{cc}
\lambda_0-\xi\alpha^2& -\xi\alpha\beta\\
-\xi\alpha\beta& \lambda_{N-1}-\xi\beta^2
\end{array}
\right],
\end{equation}
and satisfy the equation
\begin{equation}\label{eq34}
\det(\overline{\mathbf{\Theta}}-\theta\mathbf{I})=0.
\end{equation}
The quadratic form of \eqref{eq34} is expressed by
\begin{IEEEeqnarray}{ll}\label{eq35}
&\theta^2-(\lambda_0+\lambda_{N-1}-\xi)\theta\nonumber\\
&\quad\quad\quad\quad+\lambda_0\lambda_{N-1}-\xi(\alpha^2\lambda_{N-1}+\beta^2\lambda_0)=0.
\end{IEEEeqnarray}
Re-arranging \eqref{eq35} as the following form
\begin{equation}\label{eq36}
\xi=\frac{(\theta-\lambda_0)(\theta-\lambda_{N-1})}{\alpha^2\lambda_{N-1}+\beta^2\lambda_0-\theta}.
\end{equation}

To achieve the lower bound in \eqref{eq30}, we expect one eigenvalue (i.e., $\theta_0$) to meet the condition 
\begin{equation}\label{eq37}
\theta_0=\lambda_1,
\end{equation}
while the other (i.e., $\theta_{N-1}$) fulfills the condition
\begin{IEEEeqnarray}{ll}\label{eq38}
&|\theta_{N-1}|\geq\lambda_{N-1},\\
&|\theta_{N-1}|\leq\lambda_{1}.\label{eq39}
\end{IEEEeqnarray}
Meeting the condition in \eqref{eq38} is relatively straightforward. 
Therefore, our primary focus in this study will revolve around satisfying conditions \eqref{eq37} and \eqref{eq39}.

Applying \eqref{eq37} into \eqref{eq36} yields
\begin{equation}\label{eq40}
\xi=\frac{(\lambda_1-\lambda_0)(\lambda_1-\lambda_{N-1})}{\alpha^2\lambda_{N-1}+\beta^2\lambda_0-\lambda_1}.
\end{equation}
Furthermore, the condition \eqref{eq38} is equivalent to: $\theta_{N-1}\leq-\lambda_{N-2}$.
It is trivial to justify that $\xi$ is a monotonically decreasing function of $\theta$ within the range of $\theta\in(-\infty, 0)$.
Therefore, a sufficient condition to satisfy \eqref{eq38} is
\begin{IEEEeqnarray}{ll}\label{eq41}
\xi&\geq\frac{(\lambda_{N-2}+\lambda_0)(\lambda_{N-2}+\lambda_{N-1})}{\alpha^2\lambda_{N-1}+\beta^2\lambda_0+\lambda_{N-2}}\\
&>\frac{(\lambda_{N-2}+\lambda_0)(\lambda_{N-2}+\lambda_{N-1})}{\alpha^2\lambda_{0}+\beta^2\lambda_0+\lambda_{N-2}}\label{eq42}\\
&>\lambda_{N-2}+\lambda_{N-1}\label{eq43}
\end{IEEEeqnarray}
where the inequality \eqref{eq42} holds due to: $\lambda_0>\lambda_{N-1}$.

Substituting \eqref{eq40} into \eqref{eq43} yields
\begin{equation}\label{eq44}
\frac{(\lambda_1-\lambda_0)(\lambda_1-\lambda_{N-1})}{\alpha^2\lambda_{N-1}+\beta^2\lambda_0-\lambda_1}
>\lambda_{N-2}+\lambda_{N-1}>0.
\end{equation}
Due to $(\lambda_1-\lambda_0)<0$ and $(\lambda_1-\lambda_{N-1})>0$, the configuration of $(\alpha, \beta)$ must adhere to:
\begin{equation}\label{eq45}
\alpha^2\lambda_{N-1}+\beta^2\lambda_0-\lambda_1<0.
\end{equation}
Considering \eqref{eq45}, solving \eqref{eq44} yields:
\begin{equation}
\beta^2>\left(\frac{\lambda_1-\lambda_{N-1}}{\lambda_0-\lambda_{N-1}}\right)
\left(\frac{\lambda_1-\lambda_0}{\lambda_{N-2}+\lambda_{N-1}}+1\right).
\label{eq46}
\end{equation}
Given that the expression on the right-hand side of \eqref{eq46} is surely smaller than $1$, it is always possible to find a value for $\beta$ that satisfies this inequality.
Finally, the feasibility analysis can be concluded as:
\begin{thm}\label{thm02}
To achieve the lower bound specified in \eqref{eq30}, it suffices to simultaneously satisfy the inequalities \eqref{eq43} and \eqref{eq46} for $(\xi, \beta)$. 
\end{thm}

After determining the value of $\beta$, the vector ($\mathbf{p}$) can be constructed using \eqref{eq31}, following which $\mathbf{b}$ is formed as: 
\begin{equation}\label{eq47}
\mathbf{b}=\mathbf{U}\mathbf{p}.
\end{equation}

\section{Power Iteration-Assisted Approach for The Implementation of SR-$1$R}\label{secIV}
The theoretical framework presented in Section \ref{secIII} relies on the availability of EVD.
As already discussed in Section \ref{secI}, assuming the availability of EVD contradicts the objective of employing iterative matrix inversion. 
This motivates us to propose a power iteration-assisted (PIA) approach, which can determine $(\xi, \mathbf{b})$ without need of EVD.

\subsection{The Proposed PIA Approach}
The PIA approach builds upon the principles of {\em Theorem \ref{thm02}}, incorporating certain relaxations and approximations.
Our key conclusion is stated in the following theorem. 
\begin{thm}\label{thm03}
Consider an ill-conditioned matrix $\mathbf{A}$ with its eigenvalues satisfying $\lambda_0\gg\lambda_{N-2}$ and $\lambda_1\gg\lambda_{N-2}$.
A sufficient condition for $\kappa(\mathbf{\Theta})$ to achieve the lower bound specified in \eqref{eq30} is: 
\begin{equation}\label{eq48}
\xi=\lambda_0;
\end{equation}
\begin{equation}\label{eq49}
\frac{\lambda_{N-2}}{\lambda_0}\ll\beta<\frac{\lambda_{1}}{\lambda_0}.
\end{equation}
\end{thm}
\begin{proof}
{\em Theorem \ref{thm02}} indicates that $\xi$ should fulfill the condition \eqref{eq43}.
Given $\lambda_0\gg\lambda_{N-2}$ and $\lambda_{N-2}>\lambda_{N-1}$, setting $\xi$ as specified in \eqref{eq48} is sufficient to guarantee the fulfillment of condition \eqref{eq43}.
Then, we plug \eqref{eq48} into \eqref{eq35} and obtain
\begin{equation}\label{eq50}
x^2-\lambda_{N-1}x+\beta^2\lambda_0\lambda_{N-1}-\beta^2\lambda_0^2=0,
\end{equation}
where the notation $x$ is utilized in place of $\theta$ to prevent potential confusion.

Solving \eqref{eq50} gives two roots, corresponding to two eigenvalues of $\mathbf{\Theta}$:
\begin{equation}\label{eq51}
x_{1}=\frac{1}{2}\left(\lambda_{N-1}+\sqrt{\lambda_{N-1}^2+\Delta^2}\right),
\end{equation}
\begin{equation}\label{eq52}
x_{2}=\frac{1}{2}\left(\lambda_{N-1}-\sqrt{\lambda_{N-1}^2+\Delta^2}\right),
\end{equation}
where $\Delta$ is expressed as:
\begin{equation}\label{eq53}
\Delta^2=4\beta^2\lambda_0(\lambda_0-\lambda_{N-1}).
\end{equation}
From \eqref{eq51} and \eqref{eq52}, it is straightforward to see: $|x_{2}|<x_1$.
Given the condition \eqref{eq49}, $\lambda_{N-1}$ in \eqref{eq51}-\eqref{eq53} is negligibly small and thus omitted. 
Then, \eqref{eq51} and \eqref{eq52} can be approximately written as:
\begin{equation}\label{eq54}
x_1\approx\beta\lambda_0, ~ x_2\approx-\beta\lambda_0.
\end{equation}
Applying \eqref{eq49} into \eqref{eq54}, the following relationship is evident
\begin{equation}\label{eq55}
\lambda_{N-2}\ll|x_2|<x_1<\lambda_1.
\end{equation}
In this case, $\lambda_1$ and $\lambda_{N-2}$ become the maximum and minimum singular values of $\mathbf{\Theta}$, respectively. 
This immediately leads to the conclusion: $\kappa(\mathbf{R})=\kappa(\mathbf{\Theta})=\lambda_1/\lambda_{N-2}$.
{\em Theorem \ref{thm03}} is therefore proved.
\end{proof}

{\em Theorem \ref{thm03}} is appealing mainly for two reasons: {\em 1)} $\xi$ becomes exclusively reliant on $\lambda_0$. 
Given that $\lambda_0$ stands as the largest eigenvalue of $\mathbf{A}$, it can be readily acquired using the power iteration algorithm specified in {\bf Algorithm~2}, eliminating the necessity for EVD computations;
{\em 2)} the condition \eqref{eq49} offers sufficient flexibility in determining $\beta$.
For instance, $\beta$ can be specified as follows:
\begin{IEEEeqnarray}{ll}\label{eq56}
\beta&\triangleq\frac{\mathrm{trace}(\mathbf{A})-\lambda_0}{(N-1)\lambda_0}\\
&=\frac{\frac{1}{N-1}\sum_{n=1}^{N-1}\lambda_n}{\lambda_0}.\label{eq57}
\end{IEEEeqnarray}
The numerator of \eqref{eq57} is the average over eigenvalues $(\lambda_1, \lambda_2, ..., \lambda_{N-1})$. 
{It is trivial to justify that $\beta$  satisfies the condition \eqref{eq49}}.
Notably, $\beta$ defined in this form only requires the estimation of $\lambda_0$.

The last step is to form $\mathbf{b}$. 
Applying \eqref{eq31} into \eqref{eq47} results in
\begin{equation}\label{eq58}
\mathbf{b}=\alpha\mathbf{u}_0+\beta\mathbf{u}_{N-1},~\alpha^2+\beta^2=1,
\end{equation}
where $\mathbf{u}_0$, $\mathbf{u}_{N-1}$ are the $0^{th}$ and $(N-1)^{th}$ column of $\mathbf{U}$, respectively.
Since $\mathbf{u}_0$ is the eigenvector corresponding to the largest eigenvalue $\lambda_0$, it can also be determined through the power iteration algorithm.

Determining $\mathbf{u}_{N-1}$, the eigenvector corresponding to the smallest eigenvalue $\lambda_{N-1}$, poses some difficulty. 
The most commonly used algorithm is known as the inverse power iteration \cite{Arablouei2015}, which applies power iteration on $\mathbf{A}^{-1}$ as $\mathbf{u}_{N-1}$ corresponds to the largest eigenvalue of $\mathbf{A}^{-1}$. 
However, our primary objective revolves around computing $\mathbf{A}^{-1}$, making its assumed availability impractical.

{
\begin{algorithm}[t]
\caption{Pseudo code of the PIA for the SR-$1$R}\label{algo1}
\begin{algorithmic}[1]\label{agthm1}
\renewcommand{\algorithmicrequire}{\textbf{Input:}}
\renewcommand{\algorithmicensure}{\textbf{Output:}}
\REQUIRE Positive-definite Hermitian matrix $\mathbf{A}$;
\ENSURE  $\mathbf{A}^{-1}$;
\STATE Feed $\mathbf{A}$ into the power iteration function ({\bf Algorithm 2}) to acquire the estimate of $\lambda_0$ (denoted as $\hat{\lambda}_{0}$) and the estimate of $\mathbf{u}_0$ (denoted as $\hat{\mathbf{u}}_0$);
\STATE Compute $\xi$ and $\beta$ using \eqref{eq48} and \eqref{eq56}, respectively.
\STATE Construct the matrix $\mathbf{\Phi}$ in form of \eqref{eq60};
\STATE Feed $\mathbf{\Phi}$ into {\bf Algorithm 2} to acquire the estimate of $\mathbf{u}_{N-1}$ (denoted as $\hat{\mathbf{u}}_{N-1}$);
\STATE Compute the vector $\mathbf{b}$ via \eqref{eq58};
\STATE Form the matrix $\mathbf{R}$ via \eqref{eqn01} and compute $\mathbf{R}^{-1}$ through the Schulz method specified in \eqref{eq09};
\STATE Compute $\mathbf{A}^{-1}$ via the Sherman-Morrison transform \eqref{eqn02};
\RETURN $\mathbf{A}^{-1}$. 
\end{algorithmic} 
\end{algorithm}

}

The other commonly used method is referred to the shifted power iteration, which applies the power iteration algorithm on 
\begin{equation}\label{eq59}
\mathbf{\Phi}=\mathbf{A}+\gamma\mathbf{I}.
\end{equation}
The key is to find a suitable parameter $\gamma$ such that $\mathbf{u}_0$ corresponds to the maximum eigenvalue of $\mathbf{\Phi}$.
In this paper, we redefine the shifted power iteration as
\begin{equation}\label{eq60}
\mathbf{\Phi}=\gamma\mathbf{I}-\mathbf{A},
\end{equation}
and specify $\gamma$ as
\begin{equation}\label{eq61}
\gamma=\mathrm{trace}(\mathbf{A}).
\end{equation}
By this means, eigenvalues of $\mathbf{\Phi}$ are expressed as
\begin{equation}\label{eq62}
\phi_n=\sum_{k=0}^{N-1}\lambda_k, ~k\neq N-1-n.
\end{equation}
The largest eigenvalue of $\mathbf{\Phi}$ is $\phi_{N-1}$, which is corresponding to the eigenvector $\mathbf{u}_{N-1}$.
Then, applying the power iteration algorithm ({\bf Algorithm 2}) on $\mathbf{\Phi}$ yields the estimate of $\mathbf{u}_{N-1}$.

Finally, the proposed PIA approach for SR-$1$R implementation is summarized in {\bf Algorithm 1}.

\subsection{Convergence of The PIA Approach}\label{secIVb}
The PIA approach heavily depends on employing the power iteration algorithm to derive essential parameters including $\lambda_0$, $\mathbf{u}_0$, and $\mathbf{u}_{N-1}$. 
Introducing pseudo code that demonstrates the power iteration algorithm could offer valuable insights to readers interested in understanding our performance analysis.
To this end, we provide the power iteration algorithm in  ({\bf Algorithm 2}), where the matrix $\mathbf{A}$ is employed as a showcase. 
Nevertheless, $\mathbf{A}$ can be easily replaced by any Gram matrix such as $\mathbf{\Phi}$.

\begin{algorithm}[t]
\caption{Pseudo code of the power iteration algorithm}\label{algo2}
\begin{algorithmic}[1]\label{agthm2}
\renewcommand{\algorithmicrequire}{\textbf{Input:}}
\renewcommand{\algorithmicensure}{\textbf{Output:}}
\REQUIRE Matrix ($\mathbf{A}$), Tolerance ($\eta$);
\ENSURE  The largest eigenvalue ($\lambda_0$), the eigenvector $\mathbf{u}_0$;
\STATE Initialize a random vector $\mathbf{u}$ and $\lambda_{\textsc{old}}=\|\mathbf{u}\|$;
\STATE Normalize $\mathbf{u}$: $\mathbf{u}=\mathbf{u}/\|\mathbf{u}\|$;
\STATE Compute $\mathbf{v}=\mathbf{A}\mathbf{u}$;
\STATE Compute $\lambda=\|\mathbf{v}\|$;
\STATE Let $\mathbf{u}=\mathbf{v}/\|\mathbf{v}\|$;
\STATE If $|\lambda-\lambda_{\textsc{old}}|<\eta$ then break;
\STATE Let $\lambda_{\textsc{old}}=\lambda$ and goto step $3$;
\RETURN $\lambda_0=\lambda$, $\mathbf{u}_0=\mathbf{u}$. 
\end{algorithmic} 
\end{algorithm}

The convergence behavior of the power iteration algorithm has been studied in the literature \cite{Park2023} as:
\begin{equation}\label{eq63}
\lambda\rightarrow\lambda_0+\underbrace{\sum_{n=1}^{N-1}\left(\frac{\lambda_n}{\lambda_0}\right)^ic_n}_{\triangleq\widetilde{\lambda}(i)},
\end{equation}
where $i$ is the number of power iterations, and $c_n, \forall n,$ are coefficients related to both eigenvectors of $\mathbf{A}$ and the initial vector $\mathbf{u}$. 
For ill-conditioned systems where $\lambda_0$ is dominating (e.g., $\lambda_0\gg\lambda_1$), 
the term $\widetilde{\lambda}(i)$ in \eqref{eq63} becomes negligibly small only after a couple of iterations. 
This is mostly the case for ELAA-MIMO systems.


However, for the matrix $\mathbf{\Phi}$ specified in \eqref{eq60}, the convergence behavior becomes
\begin{equation}\label{eq64}
\phi\rightarrow\phi_0+\underbrace{\sum_{n=1}^{N-1}\left(\frac{\phi_n}{\phi_0}\right)^ic'_n}_{\triangleq\widetilde{\lambda}(i)},
\end{equation}
where $c_n', \forall n,$ are corresponding coefficients. 
When the values of $c_n', \forall n,$ exhibit comparability among themselves, the dominant factor influencing convergence is expressed by
\begin{equation}\label{eq65}
\frac{\phi_1}{\phi_0}=\frac{\lambda_{N-1}+\sum_{k=0}^{N-3}\lambda_k}{\lambda_{N-2}+\sum_{k=0}^{N-3}\lambda_k}.
\end{equation}
Given that the eigenvalues $(\lambda_{N-1}, \lambda_{N-2})$ are significantly smaller compared to the sum of the remaining eigenvalues, the ratio specified in \eqref{eq65} is close to $1$. 
Consequently, this leads to notably slow convergence when determining the eigenvector $\mathbf{u}_{N-1}$.
In order to speed up the convergence, an enhanced PIA approach is proposed in Section \ref{4c}.

%

\subsection{Enhanced PIA Approach}\label{4c}
The opportunity for convergence acceleration arises from the coefficients $c_n', \forall n$, in \eqref{eq64}.
Following the principle of power iteration \cite{Park2023}, the coefficients $c_n', \forall n,$ are related to the matrix $\mathbf{\Phi}$ and the initial vector $\mathbf{u}$.
Given that $\mathbf{\Phi}$ and $\mathbf{u}$ are randomly generated and independent of each other,  
$\mathbf{c}'\triangleq(c_1', c_2', ..., c_{N-1}')^T$ can be viewed as a randomly generated sequence normalized in the sequence norm.

Define $\matc{\phi}(i)\triangleq(\phi_1^i, \phi_2^i, ..., \phi_{N-1}^i)^T/\phi_0^i$ and represent the term $\widetilde{\lambda}(i)$ in \eqref{eq64} as the following vector form
\begin{equation}\label{eq66}
\widetilde{\lambda}(i)=\matc{\phi}^T(i)\mathbf{c}'.
\end{equation}
Given the matrix $\mathbf{\Phi}$, $\mathbf{c}'$ is solely dependent on $\mathbf{u}$. 
When producing $L$ independent generations of $\mathbf{u}$ denoted as $\mathbf{u}_0, \mathbf{u}_1, ..., \mathbf{u}_{L-1}$, their corresponding vectors $\mathbf{c}_0', \mathbf{c}_1', ..., \mathbf{c}_{L-1}'$ are generated.
Then, the enhanced PIA (e-PIA) approach aims at solving the following objective function
\begin{equation}\label{eq67}
l^\star=\underset{l}{\arg\min}|\matc{\phi}^T(i)\mathbf{c}'_l|, ~_{l=0, ..., L-1}.
\end{equation}
This optimization problem can be easily managed via the power iteration algorithm specified in {\bf Algorithm 2}.
For various generations of $\mathbf{u}$ (or equivalently $\mathbf{c}'$), the power iteration algorithm can run in parallel. 
The power iteration process related to $l^\star$ will be the first to reach the saturation point $\eta$ as specified in Step 6 of {\bf Algorithm 2}, and consequently have the convergence accelerated. 

Generally, it is mathematically challenging to quantify the gain of convergence acceleration. 
However, we can find a faster convergence speed that can be achieved when $\mathbf{c}'\triangleq(0, 0, ..., 1)^T$.
In this extreme case, the dominating factor influencing convergence becomes
\begin{equation}\label{eq68}
\frac{\phi_{N-1}}{\phi_0}=\frac{\lambda_{N-1}+\sum_{k=1}^{N-2}\lambda_k}{\lambda_{0}+\sum_{k=1}^{N-2}\lambda_k}.
\end{equation}
For ill-conditioned systems where $\lambda_0$ is the dominant eigenvalue, this factor is significantly smaller than the one outlined in \eqref{eq65}.

\subsection{Scalability Analysis for SR-1R}\label{secIVD}
In this subsection, we analyze the scalability of iterative linear precoding with the SR-$1$R in terms of both the overall complexity and the algorithm-depth.	
Then, we compare the SR-$1$R to the preconditioning baselines.
The complexity analysis reflects the computational cost of the SR-$1$R; while the algorithm-depth reflects the minimum computing steps (i.e., the shortest processing time) when employing parallel computing \cite{parallelComput}.
This measure is particularly relevant because compatibility with parallel computing is a key advantage of iterative matrix inversion methods \cite{10278625}.
It will be shown that both the PIA/e-PIA algorithms exhibit favorable compatibility with parallel computing.

Before discussing \textbf{Algorithm~\ref{agthm1}}, we present some basic conclusions of the algorithm-depth when the computing units are assumed sufficient: The multiplication of a matrix/vector by a $N\times1$ vector yields a depth of $\log_{2}N$; the summation of $N$ scalars yields a depth of $\log_{2}N$ (e.g., calculating the trace of a $N\times N$ matrix or the norm of a $N\times1$ vector) \cite{parallelComput}.
Then, we analyze the PIA approach (\textbf{Algorithm~\ref{agthm1}}) step by step.

\begin{itemize}
\item Step $1$. Each iteration of \textbf{Algorithm~\ref{agthm2}} calculates a matrix-vector multiplication and a norm of a vector.
Denote the iteration number of \textbf{Algorithm~\ref{agthm2}} by $\tau_0$.
The complexity is $\tau_0N^2$, while the algorithm-depth is $\tau_0\log_{2}N$.
\item Step $2$. $\xi$ is the direct output of \textbf{Algorithm~\ref{agthm2}} in Step~$1$.
The calculation of $\beta$ requires $\mathrm{Tr}(\mathbf{A})$.
This has $\log_{2}N$ algorithm-depth and $N$ complexity.
\item Step $3$. With $\mathrm{Tr}(\mathbf{A})$ already available in Step~$2$, the calculation of $\mathbf{\Phi}$ has $N$ complexity and negligible depth.
\item Step $4$. Denote the iteration number of \textbf{Algorithm~\ref{agthm2}} by $\tau_{N-1}$.
Similar to Step~$1$, the complexity is $\tau_{N-1}N^2$, while the algorithm-depth is $\tau_{N-1}\log_{2}N$.
\item Step $5$. The summation of two vectors has $N$ complexity and negligible depth.
\item Step $6$. This step belongs to the iterative matrix inversion. 
Each iteration costs two matrix-matrix multiplications and a matrix-matrix deduction.
When the iteration nmber is $i$, this yields the complexity of $(2N^3+N^2)i$ and the algorithm-depth of $2i\log_{2}N$.
\item Step $7$. The second term of the Sherman-Morrison transform (recall \eqref{eqn02}) can be decomposed into three matrix-vector multiplications, a vector outer product and a vector inner product.
There is also a matrix-matrix deduction between the first and the second term.
Note the matrix-vector multiplications and the inner product can be computed in parallel; while both the outer product and the deduction has negligible depth.
This yields the algorithm-depth of $\log_2 N$ and the complexity of $5N^2+N$.
\end{itemize}

Apart from the complexity and algorithm-depth in \textbf{Algorithm~\ref{agthm1}}, there are some extra steps in iterative linear precoding.
Their complexity and algorithm-depth are described as follows.
\begin{itemize}
\item Calculation of $\mathbf{A}=\mathbf{H}\mathbf{H}^H$. 
This step has the complexity of $MN^2$ and the algorithm-depth of $\log_{2}M$.
\item Calculation of the precoded signal $\mathbf{x}=\mathbf{H}^H\mathbf{A}^{-1}\mathbf{s}$.
This involves two matrix-vector multiplications with the complexity of $N^2+MN$ and the algorithm-depth of $2\log_{2}N$.

\item Calculation of $\omega$ to ensure the inequality \eqref{eq10} (replace $\mathbf{A}$ with $\mathbf{R}$).
There are various ways to determine $\omega$.
In this paper, the Gershgorin circle theorem is employed (see \cite{Saad2003}):
\begin{equation}\label{eq78}
\omega_\mathrm{Ger}=\Big(\max_n\big(\sum_{i=0}^{N-1}|\mathbf{R}_{n}^H\mathbf{R}_{i}|\big)\Big)^{-1},
\end{equation}
where $\mathbf{R}_n$ stands for the $n^{th}$ column of $\mathbf{R}$.
The calculation of Gershgorin circle theorem involves three parts: calculation of each element $\mathbf{R}_{n}^H\mathbf{R}_{i}$ (equivalent to a matrix-matrix multiplication); summations of $|\mathbf{R}_{n}^H\mathbf{R}_{i}|$ (each summation can be computed in parallel); ordering of summation results.
The ordering has the complexity of $N\log_{2}N$ and the algorithm-depth of $\log_{2}N$ \cite{parallelComput}.
Hence, the complexity of the Gershgorin theorem is $N^3+N^2+N\log_{2}N$, and the algorithm-depth is $3\log_{2}N$.

\end{itemize}

Summing up these steps apart from Step~$6$ yields the overall algorithm depth of the SR-$1$R:
\begin{equation}\label{eq79}
(\tau_0+\tau_{N-1}+7)\log_{2}N+\log_{2}M.
\end{equation}
Step~$6$ is excluded in \eqref{eq79}, as it belongs to the iterative matrix inversion itself rather than to the preconditioning or regularization procedures.
A good regularization/preconditioning technique accelerates this process.
In Section~\ref{secVD}, we will demonstrate the iterations needed for convergence and computational latency when the iterative matrix inversion converges~(see Table~\ref{tabLatency}).

The overall computational complexity of the SR-$1$R is:
\begin{equation}
N^3+(\tau_0+\tau_{N-1}+M+7)N^2+(M+\log_{2}N+4)N.\label{eq81_rev2}
\end{equation}
We observe that this complexity level is at the same level as that of direct inversion, which is $\mathcal{O}(N^3)$ \cite{Wang2022Gower}. 
However, as emphasized in the introduction, this additional complexity is a common trade-off to enhance compatibility with parallel computing architectures \cite{Borkar2011,Kruskal1990}.

When employing the e-PIA, \textbf{Algorithm~\ref{agthm1}} is performed in parallel for $L$ times.
Therefore, the complexity level of the e-PIA is $\mathcal{O}(LN^3)$; while the algorithm-depth level remains the same as the PIA.

The preconditioning baselines usually involves the inversion of a triangular matrix, i.e., a complexity of $\mathcal{O}(N^2)$ \cite{Liu2023}.
This is done by the Gauss-Jordan elimination, and constitutes $N$ iterations \cite{JF_RB_1995}.
Hence, the algorithm-depth is $\mathcal{O}(N)$.
After taking the necessary extra steps (e.g., \eqref{eq78}) into consideration, the overall complexity of preconditioning baselines is dominated by the Gershgorin circle theorem, i.e., $\mathcal{O}(N^3)$.
While the algorithm depth is dominated by the triangular matrix inversion, i.e., $\mathcal{O}(N)$.
Hence, the SR-$1$R reduces the algorithm depth from linear level to logarithm level of $N$, while maintaining the complexity at the same level as the preconditioning baselines.

\begin{figure*}[t]	
	\centering
	\begin{minipage}[t]{0.66\textwidth}
		\centering
		\includegraphics[width=1\textwidth]{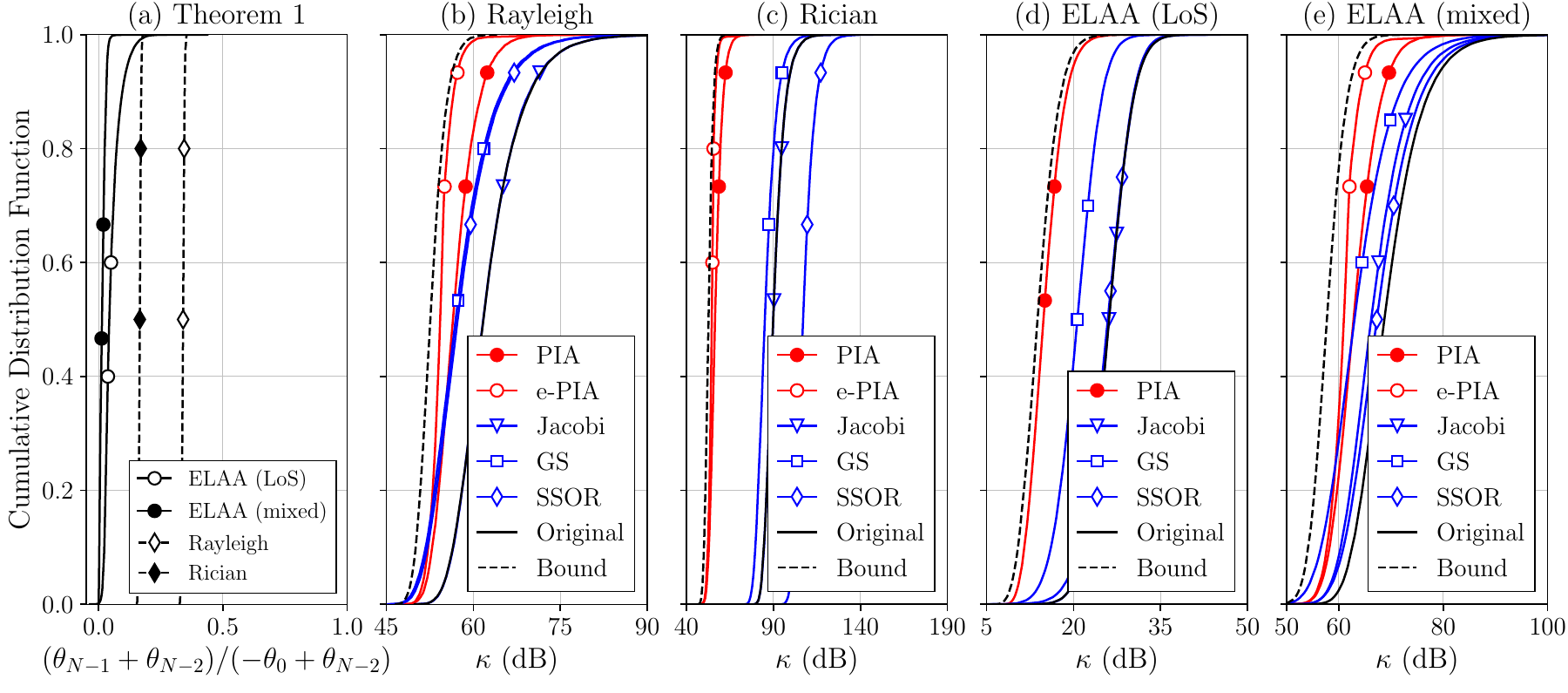}
		\vspace{-2.3em}
		\caption{Results for \textit{Experiment~1}: (a) CDF of $(\theta_{N-1}+\theta_{N-2})/(-\theta_{0}+\theta_{N-2})$ for the PIA approach; CDF of condition number compared to the preconditioning baselines in (b)~i.i.d. Rayleigh channel, (c)~i.i.d. Rician channel, (d)~LoS-dominated ELAA, (e)~mixed LoS/non-LoS ELAA.}\label{figCondPDF_stationary512}
	\end{minipage}
	\begin{minipage}[t]{0.33\textwidth}
		\centering
		\includegraphics[width=1\textwidth]{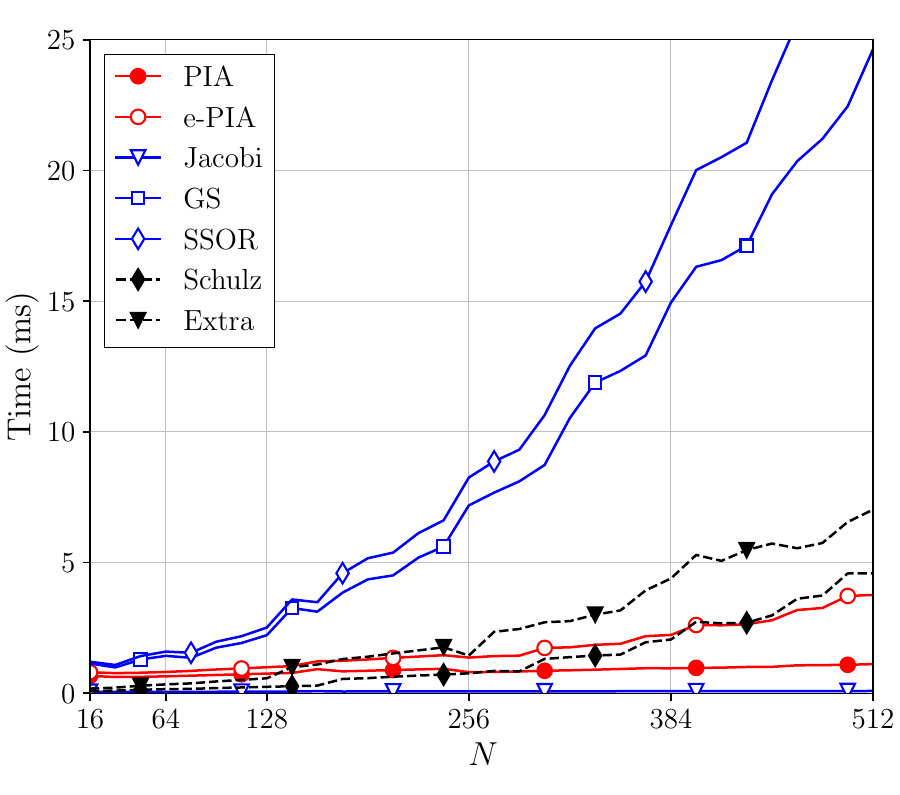}
		\vspace{-2.3em}
		\caption{Computation latency of the PIA/e-PIA approaches compared to the preconditioning baselines as $N$ increases from $16$ to $512$.}
		\label{fig61}
	\end{minipage}
	\vspace{-2.2em}
\end{figure*}

In Section~\ref{secV}, we will demonstrate the scalability of the SR-$1$R in terms of the computation latency when parallel computing is supported (see Fig.~\ref{fig61}) as well as the overall latency of iterative linear precoding when RZF performance is achieved (see Table~\ref{tabLatency}).

\section{Simulation Results and Discussion}\label{secV}

In this section, computer simulations are carried out to demonstrate the performance of both the PIA and e-PIA approaches for the SR-$1$R.
Specifically, Section~\ref{secVA} shows the implementation details of the SR-$1$R; Section~\ref{secVB} details the baselines and simulation setup; Section~\ref{secVD} shows simulation results.

%
%
%

\subsection{Implementation of the SR-$1$R}\label{secVA}
Setting the convergence threshold $\eta$ in \textbf{Algorithm~\ref{agthm2}} could be difficult in implementation.
This is because the $\eta$ could be reached due to bad initialization.
The increment of \textbf{Algorithm~\ref{agthm2}} is given by
\begin{equation}\label{eq83}
\lambda-\lambda_\textsc{old}=\sum_{n=1}^{N-1}\Big(\frac{\lambda_n}{\lambda_0}\Big)^{i-1}\Big(\frac{\lambda_n}{\lambda_0}-1\Big).
\end{equation}
\eqref{eq83} means the increment approaches zero in two cases: when $i\to\infty$, or when $(\lambda_n)/(\lambda_0)\approx1$.
When $\mathbf{u}$ is badly initialized, i.e., $\lambda_n\approx\lambda_{0}$, it is hard to determine the convergence.
This brings troubles to the PIA and e-PIA approaches to calculate  $\hat{\mathbf{u}}_{N-1}$ .

To handle this problem, we fix the iteration number of \textbf{Algorithm~\ref{agthm2}} in the implementation, i.e., $\tau_0=\tau_{N-1}=\tau$.
Moreover, to avoid picking the wrong candidate in the e-PIA, each candidate is fed into the iterative matrix inversion to produce an inversion result, denoted by $\mathbf{A}^{-1}_l$, $_{l=0,\cdots,L-1}$.
Then, we choose the inversion result that minimizes the residual error as the output of the e-PIA approach:
\begin{equation}\label{eq84}
\mathbf{A}^{-1}=\argmin_{\mathbf{A}^{-1}_l}\|\mathbf{I}-\mathbf{A}\mathbf{A}^{-1}_l\|^2_\mathrm{F}.
\end{equation}

Notably, the algorithm-depth of the e-PIA approach remains logarithmic with this implementation.
The computation of each candidate can be parallelized.
The calculation of the residual error is a summation of $N^2$ scalars, and has an algorithm-depth of $\mathcal{O}(\log_{2}N)$.
The ordering of the residual error has a depth of $\mathcal{O}(\log_{2} L)$ \cite{parallelComput}.


\subsection{Baselines and Simulation Setup}\label{secVB}

We consider two types of baselines: \textbf{1) iterative matrix inversion}, where we use the Schulz method (see \eqref{eq09})~\cite{Schulz1933}; \textbf{2) preconditioning}, where we use the Jacobi preconditioning (see \eqref{eq14New}) \cite[Eq.~(10)]{8417575}, Gauss-Seidel (GS) preconditioning (see \eqref{eq15New}) \cite[Eq. (14)]{Wang2023Hanzo}, and Symmetric successive over-relaxation (SSOR) (see \eqref{eq16New}).


We consider two types of stationary ill-conditioned channel and two types of ELAA channel in the simulation.
For stationary channels, we consider \textbf{1) Symmetric i.i.d. Rayleigh}, where the ill condition arises from the symmetric MIMO size;
\textbf{2) Symmetric i.i.d. Rayleigh}, where the ill condition also arises from the strong LoS correlation.
In most cases, we set $M = N = 512$, as iterative matrix inversion is particularly beneficial in large-scale MIMO systems.

For ELAA channels, the channel model in \cite{10295381} is adopted for the ELAA channel in appreciation of its comprehensive consideration of mixed LoS/non-LoS states, shadowing effects and path-loss.
We consider the height of users and the base station to be $1.5$~m and $10$~m, respectively.
Specifically, the considered ELAA channels include: 
\textbf{1) LoS-dominated ELAA}, where the ill condition arises from the correlation of the LoS paths.
We consider $M=64$, $N=16$ (single user), and the vertical distance from the user to the ELAA is $40$~m;
\textbf{2) Symmetric Mixed LoS/non-LoS ELAA}, where the ill condition arises from both the correlation of the LoS paths and the symmetric MIMO size.
The MIMO size is configured as: $M=N=128$ ($8$ users with $16$ antennas each), and the vertical distance from users to the ELAA is $40$~m.
The users are equally spaced along a parallel line to the base station antenna array with a maximal inter-user distance of $10$~m.

In the simulation, the wireless channel is normalized ($\|\mathbf{H}\|_\mathrm{F}^2=N$) to remove the effect of path loss in ELAA channels.
The carrier frequency is $3.5$~GHz; the modulation scheme is $256$~QAM; the power iteration number is $\tau=1$.


\subsection{Simulation Results and Discussion}\label{secVD}

\begin{figure*}[t]	
	\centering
	\begin{minipage}[t]{0.33\textwidth}
		\centering
		\includegraphics[width=1\textwidth]{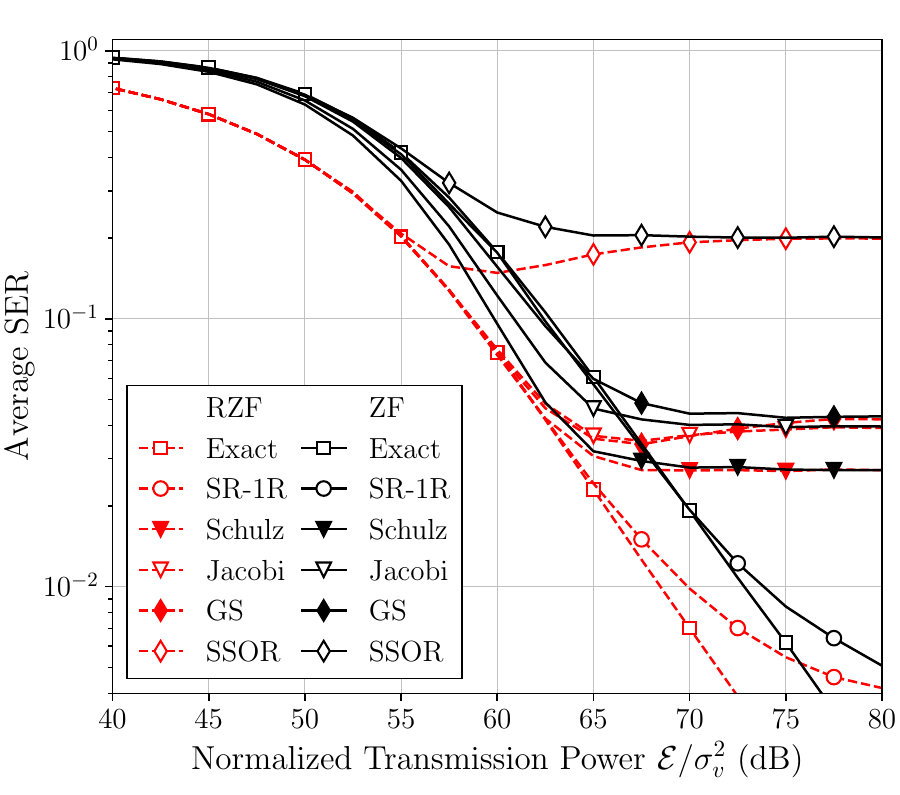}
		\vspace{-2.3em}
		\caption{Average SER of the e-PIA ($i=46$) and the baselines ($i=50$) for the RZF and ZF w.r.t. $(\mathcal{E})/(\sigma_v^2)$ in mixed LoS/non-LoS ELAA.}
		\label{fig7}
	\end{minipage}
	\begin{minipage}[t]{0.66\textwidth}
		\centering
		\includegraphics[width=1\textwidth]{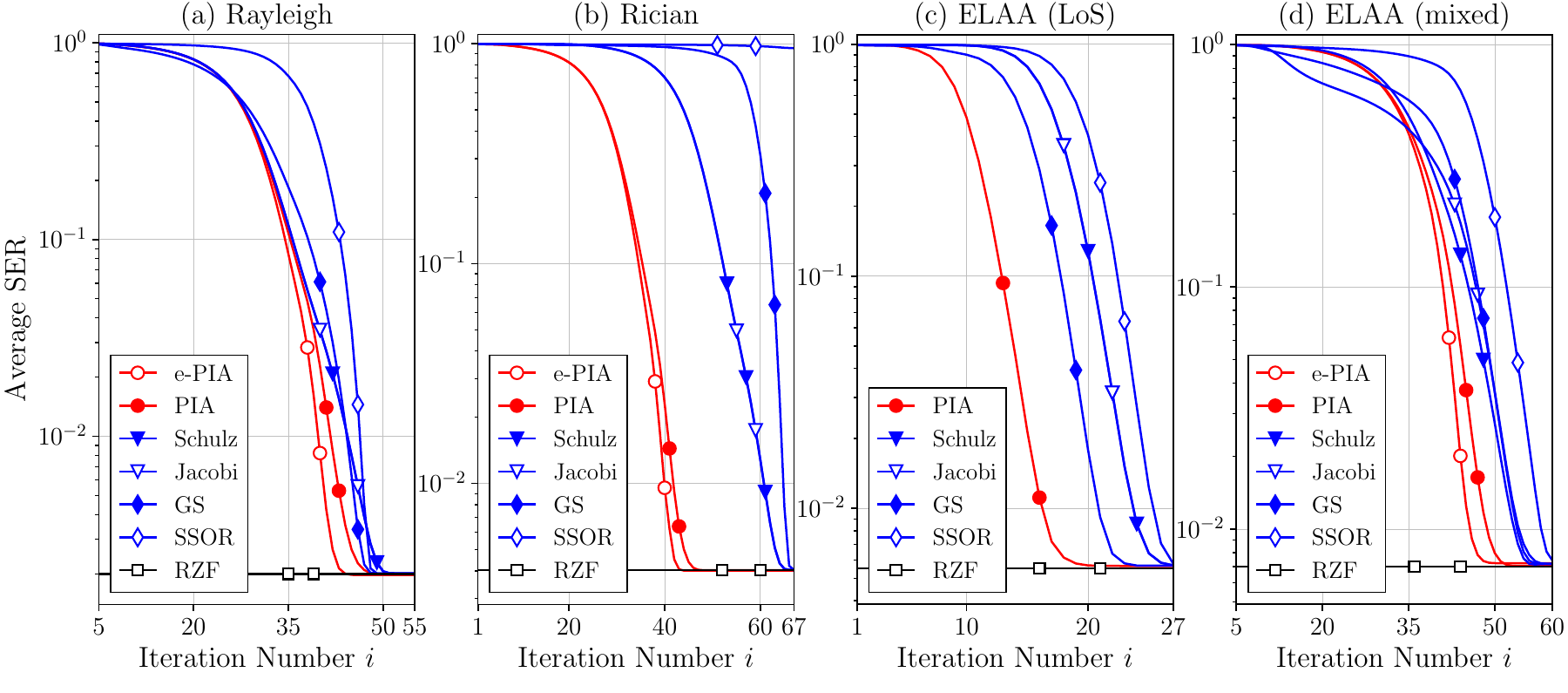}
		\vspace{-2.3em}
		\caption{Average SER of the PIA/e-PIA approaches as the iteration number increases compared to exact RZF performance in (a)~i.i.d. Rayleigh channel, (b)~i.i.d. Rician channel, (c)~LoS-dominated ELAA, (d)~mixed LoS/non-LoS ELAA.}
		\label{fig512Iter}
	\end{minipage}
	\vspace{-2.2em}
\end{figure*}

\subsubsection{Experiment 1}
This experiment aims to investigate the performance of the proposed PIA/e-PIA approaches in comparison to the theoretical expectation in \textit{Theorem~\ref{thm01}} as well as in terms of condition number in comparison to the preconditioning baselines, as demonstrated in Fig.~\ref{figCondPDF_stationary512}.
Here we consider the ZF precoding, i.e., $\mathbf{A}=\mathbf{H}\mathbf{H}^H$.
The performance using the RZF will be shown in {\textit{Experiment~3}$\sim$\textit{4}}.

The first thing of interest is the behavior of $\theta_{N-1}$: \textit{Theorem~\ref{thm01}} expects $\theta_{N-1}\in(-\theta_{0},-\theta_{N-2})$.
Since $\theta_{0}$ and $\theta_{N-2}$ change in each realization of the channel, we use the following measurement:
\begin{equation}\label{eq86}
\frac{\theta_{N-1}-(-\theta_{N-2})}{(-\theta_{0})-(-\theta_{N-2})}=\frac{\theta_{N-1}+\theta_{N-2}}{-\theta_{0}+\theta_{N-2}}.
\end{equation}
\eqref{eq86} falls in $(0,1)$ when $\theta_{N-1}\in(-\theta_{0},-\theta_{N-2})$.

Fig.~\ref{figCondPDF_stationary512}(a) demonstrates the cumulative distribution function (CDF) of \eqref{eq86}, where the solid lines stand for the ELAA channels and the dash lines for the stationary channels.
It is shown that the measurement in \eqref{eq86} falls in the expected interval.
Hence, the PIA approach successfully fulfills \textit{Theorem~\ref{thm01}}, such that $\kappa(\mathbf{R})=(\theta_{0})/(\theta_{N-2})$.

Then, we investigates the condition number improvement of the PIA/e-PIA approaches, as shown in Fig.~\ref{figCondPDF_stationary512}(b)$\sim$(e).
Specifically, the black solid line stands for the CDF of $\kappa(\mathbf{A})$ (denoted by `Original'), while the black dash line stands for the CDF of $(\lambda_{1})/(\lambda_{N-2})$ as the performance bound of the SR-$1$R (denoted by `Bound').
The red lines with circle markers stand for the proposed PIA/e-PIA approaches, while the blue lines stand for the preconditioning baselines.

We first discuss the case where the channel is stationary.
Fig.~\ref{figCondPDF_stationary512}(b) shows the CDF of the condition number in i.i.d. Rayleigh channel when $M=N=512$.
It is shown that the PIA significantly reduces the tail of the CDF.
For example, when the CDF function is $0.9$, the PIA is around $5$~dB smaller than the GS and SSOR preconditioning.
In communication systems, this indicates that the GS and SSOR will need more iterations to handle the extremely large condition number such that the communication error can be minimized in average.
In addition, the e-PIA extends this advantage to around $10$~dB.
As for the Jacobi preconditioning, it hardly changes the condition number from the original case.

When the channel is i.i.d. Rician, the improvement brought by the PIA/e-PIA is even more significant.
As shown in Fig.~\ref{figCondPDF_stationary512}(c), the PIA/e-PIA reduces the condition number for around $25$~dB.
Moreover, in both types of channels, the e-PIA is close to the performance bound. 

Fig.~\ref{figCondPDF_stationary512}(d)$\sim$(e) demonstrates the CDF of the condition number when the channel is spatially non-stationary.
When the channel is LoS-dominated, the PIA can reach the performance bound already, as shown in Fig.~\ref{figCondPDF_stationary512}(d).
In this case, the PIA improves the condition number for around $6$~dB compared to the GS preconditioning.
When the channel is mixed LoS/non-LoS, as shown in Fig.~\ref{figCondPDF_stationary512}(e), the PIA/e-PIA shows similar behavior to the case of i.i.d. Rayleigh fading, i.e., the PIA and e-PIA achieve around $5$~dB and $10$~dB improvement when the CDF function is $0.9$, respectively.

Therefore, it is concluded that the PIA/e-PIA significantly improves the condition number.
Moreover, the e-PIA substantially reduces the gap between the PIA and the performance bound when the MIMO size is symmetric.

\subsubsection{Experiment 2}

This experiment aims to investigate the scalability of the SR-$1$R by showing the computation latency, as shown in Fig.~\ref{fig61}.
The computation device is Nvidia RTX A6000.
The results are averaged based on $5,000$ implementations.
Notably, substantial efforts have been made to accelerate matrix inversion on GPU platforms (e.g., \cite{Yu2015, Wang2020}). 
Here, the matrix partitioning is employed to accelerate the computation of GS and SSOR \cite[\textbf{Theorem~1}]{Wang2020}.
It is shown that the PIA roughly increases from around $0.7$~ms to $1.3$ ms as $N$ increases from $16$ to $512$.
The computation latency of the e-PIA is higher than the PIA, e.g., $3.8$~ms when $N=512$, yet still at the same level.
This coincides with our scalability analysis.
In conclusion, both the PIA/e-PIA show outstanding scalability.
In comparison, the GS and SSOR reach $25$~ms and $31$~ms when $N=512$, respectively.

Fig.~\ref{fig61} also includes the latency of one iteration for the Schulz method and the extra latency to calculate the linear precoder, as shown by the black dash lines.
They will be used to compared the overall latency for iterative linear precoding later in \textit{Experiment~4}.

\subsubsection{Experiment 3}
	
This experiment aims to investigate the behavior of iterative linear precoding with the RZF compared to the ZF.
We choose the case of mixed LoS/non-LoS ELAA as representative. 
In other cases, the phenomenon observed is similar.
To facilitate the analysis, define the SNR as the ratio of the transmission power $\mathcal{E}$ over the noise power $\sigma_v^2$, i.e., $(\mathcal{E})/(\sigma_v^2)$.
This is also referred to as the normalized transmission power.

It is expected that the improvement of the RZF on $\kappa(\mathbf{A})$ decreases as the SNR increases, because the regularization term is inversely proportional to the SNR (recall \eqref{eqn10}).
One may suggest fixing the regularization term of the RZF, but this deprives the RZF of its optimality as a linear precoder.

As shown in {Fig.~\ref{fig7}}, the red dash lines and the black solid lines stand for the case of RZF and ZF, respectively.
The RZF improves the average SER of iterative linear precoding for around $4$~dB when $(\mathcal{E})/(\sigma_v^2)=50$~dB, but gradually converges to the ZF performance.
This coincides with our expectation.

In addition, two interesting phenomenons are observed.
The first is that the preconditioning baselines for ZF sometimes outperforms the exact ZF precoding.
This is because of the estimation error of $\mathbf{A}^{-1}$, and this error plays a similar role to the regularization term of the RZF.
In contrary, the e-PIA estimates $\mathbf{A}^{-1}$ more accurately, and coincides with the exact ZF precoding when $(\mathcal{E})/(\sigma_v^2)\leq70$~dB.
The second is that the iterative linear precoding for RZF may sometimes outperform its error floor, e.g., the SSOR when  $(\mathcal{E})/(\sigma_v^2)=57.5$~dB.
The is because of a trade-off here: $\kappa(\mathbf{A})$ is better when $(\mathcal{E})/(\sigma_v^2)$ is smaller, but the SNR is higher when $(\mathcal{E})/(\sigma_v^2)$ is larger.
The trade-off yields an optimal SER when $(\mathcal{E})/(\sigma_v^2)$ is moderate.
Nevertheless, this optimal SER is close to the error floor, and does not fundamentally change the performance.

Overall, the RZF improves the SER, but does not change the behavior of iterative linear precoding
In \textit{Experiment~4}, the RZF will be employed to for further investigation.

\subsubsection{Experiment 4}

This experiment aims to investigate the iteration number of the PIA/e-PIA to achieve RZF performance in terms of average SER, as shown in Fig.~\ref{fig512Iter}(a)$\sim$(d).
Moreover, we estimate the overall latency for iterative linear precoding based on the iteration number and the results of Fig.~\ref{fig61}, as shown in Table~\ref{tabLatency}.
Since the SER of interest in uncoded systems is usually $10^{-2}$ \cite{Wang2022c}, we choose the transmission power where the RZF performance is slightly lower than $10^{-2}$.
It is worth mentioning that, for comparison to the SR-$1$R, the preconditioning baselines are combined with the Schulz method rather than the Neumann series typically used in the literature. 
This can lead to different behavior from that reported in prior studies, e.g., \cite{Wang2022Gower,8417575,Lee2020,7399337,Saad2003}.

We first discuss the performance when the channel is stationary.
Fig.~\ref{fig512Iter}(a) and Fig.~\ref{fig512Iter}(b) shows the case of $M=N=512$, where the SNR is $(\mathcal{E})/(\sigma_v^2)=65$~dB for the i.i.d. Rayleigh channel and $(\mathcal{E})/(\sigma_v^2)=70$~dB for the i.i.d. Rician channel.
It is shown that the PIA and e-PIA converges to RZF performance when the iteration number is $i=43$ and $i=46$, respectively.
In comparison, the smallest iteration number is $i=48$ of GS preconditioning in Rayleigh channel and $i=65$ of the Schulz method alone in Rician channel.
This yields a reduction of iteration number of $4\%\sim29\%$ for the PIA and $10\%\sim35\%$ for the e-PIA, and demonstrates the advantage of both the PIA and e-PIA of reducing the convergence iteration number, coinciding with our discussion in \textit{Experiment~1}.

Notably, GS preconditioning requires more iteration number to converge in Rician channel, although it showed advantage in average condition number in Fig.~\ref{figCondPDF_stationary512}(c).
This is because when the channel is extremely ill-conditioned, GS will has similar behavior as the SSOR (i.e., making the condition number worse).
This leads to burst errors when the channel is extremely ill-conditioned, and makes the SER worse. 

When the channel is spatially non-stationary, the PIA/e-PIA are still able to reduce the iteration number, as shown in Fig.~\ref{fig512Iter}(c)$\sim$(d).
The SNR for the LoS-dominated ELAA is $(\mathcal{E})/(\sigma_v^2)=42.5$~dB and for the mixed LoS/NLoS is $(\mathcal{E})/(\sigma_v^2)=70$~dB.
In LoS-dominated ELAA, using the PIA is already close to the performance bound of SR-$1$R, as shown in Fig.~\ref{fig512Iter}(c).
Hence, we only use the PIA and it converges to RZF performance when $i=19$.
In comparison, the GS preconditioning achieves RZF performance when $i=23$.
In mixed LoS/non-LoS ELAA, the PIA and e-PIA converges to RZF performance when $i=51$ and $i=48$, respectively.
In comparison, the Schulz method alone achieves RZF performance when $i=56$.

\begin{table}[t]
	\center
	\caption{Computation Latency of \\Iterative Linear Precoding}
	\label{tabLatency}
	\begin{tabular}{c||c|c|c|c}
		\hline
		& e-PIA & PIA & Schulz & GS \\
		\hline
		Rayleigh, $N=512$ & $208$~ms & $219$~ms & $241$~ms & $252$~ms \\
		\hline
		Rician, $N=512$ & $208$~ms & $219$~ms & $305$~ms & $339$~ms \\
		\hline
		ELAA (LoS) & - & $2.45$~ms & $2.37$~ms & $3.08$~ms \\
		\hline
		ELAA (mixed) & $13.6$~ms & $14.0$~ms & $14.4$~ms & $16.8$~ms \\
		\hline
	\end{tabular}
	\vspace{-2.2em}
\end{table}

With the latency measurement in Fig.~\ref{fig61} and the iteration number in Fig.~\ref{fig512Iter}, we are able to calculate the overall computation latency of iterative linear precoding, as shown in Table~\ref{tabLatency}.
We only present the Schulz method alone and the GS preconditioning in Table~\ref{tabLatency} since they need the least number of iterations to converge to RZF performance in Fig.~\ref{fig512Iter}.
The Jacobi preconditioning usually requires the same iteration as the Schulz method alone, while the SSOR requires more iterations than the GS.

Table~\ref{tabLatency} shows the calculated latency.
The first thing to be noticed is that the Schulz method is faster than the GS preconditioning in general.
This is because of the long computation latency of the GS preconditioning.
Compared to the Schulz method, the PIA achieves $9.1\%\sim28\%$ reduction of latency compared to using the Schulz method alone, while the e-PIA extends this reduction to $13.7\%\sim32\%$.
This demonstrates that the PIA/e-PIA can transform its advantage of fast convergence into shorter computation latency.

In LoS-dominated ELAA ($N=16$), the PIA is slightly slower than using the Schulz method alone.
This is mainly because the computation of \textbf{Algorithm~\ref{agthm1}} is executed in serial in MATLAB, resulting a long computation latency even when the MIMO size is small.
Nevertheless, when the channel statistics is known by the transmitter, $\xi$ and $\beta$ can be replaced by a constant \cite{10278625}.
This provides the feasibility to further reduce the computation latency of the PIA/e-PIA.

It is worth noting that the latency results reported in Table~\ref{tabLatency} may appear less favorable compared to direct matrix inversion when evaluated on current hardware platforms. 
This is primarily because existing parallel computing implementations still incurs a latency above linear scalability for large matrices, as demonstrated by the `Schulz' line in Fig.~\ref{fig61}. 
Nonetheless, this limitation is expected to diminish with the continued advancement of parallel processing technologies \cite{Zhang2016,Norimoto2023}. 
In this work, we focus on the comparative performance of iterative methods.

\section{Conclusion}\label{secVI}
In this paper, we have proposed to regularize the condition number of the channel's Gram matrix with a symmetric rank-$1$ matrix in {MIMO systems}, termed the SR-$1$R.
The SR-$1$R can align the largest/smallest eigenvalues to their closest neighbor when the rank-$1$ matrix is formed by their corresponding eigenvectors.
To enable the SR-$1$R in real-time signal processing, we have proposed the PIA approach, where the power iteration is employed to estimate these eigenvectors. 
Moreover, the e-PIA approach has been proposed to enhance the estimation of the eigenvector for the smallest eigenvalue.
Both the PIA/e-PIA approaches have good compatibility to parallel computing with logarithmic algorithm-depth.

Simulation results have demonstrated that the PIA approach significantly improves the iteration number of iterative linear precoding for {$4\%\sim29\%$} in stationary channels and ELAA channels compared to the preconditioning techniques and linear system inverse; and the e-PIA approach extends the improvement to {$10\%\sim35\%$}.
{Moreover, both the PIA/e-PIA approaches have demonstrated good scalability and parallel computing compatibility, and the reduction of iteration number almost directly translates into lower computation latency.}







\ifCLASSOPTIONcaptionsoff
\newpage
\fi

\bibliographystyle{myIEEEtran}
\bibliography{URLLC,Bib_Else,Books_and_Standards,NLP_Downlink,GroupPaper,IterativePrecoding}


~\\~\\~\\~\\~\\~\\~\\~\\~\\~\\~\\~\\~\\~\\~\\~\\~\\~\\~\\~\\~\\~\\~\\~\\~\\~\\~\\~\\~\\~\\~\\~\\~\\~\\~\\~\\

\newpage
\appendix[Experiment of Rank-$K$ Regularization]

While the SR-$1$R has demonstrated great performance improvement, its performance is limited by $\lambda_{1}$ and $\lambda_{N-2}$.
It would be interesting to know the potential to go beyond this limit.
Here we present an example using a rank-$K$ ($K>1$) matrix to regularize $\mathbf{A}$.
In this case, the regularized matrix in \eqref{eqn01} becomes:
\begin{equation}
	\mathbf{R}=\mathbf{A}-\mathbf{B}\mathbf{\Xi}\mathbf{B}^H,
\end{equation}
where $\mathbf{B}\in\mathbb{C}^{N\times K}$, and $\mathbf{\Xi}\in\mathbb{C}^{K\times K}$ is a diagonal matrix.
In this case, $\mathbf{A}^{-1}$ can be recovered from $\mathbf{R}^{-1}$ with the Woodbury formula \cite{Petersen2012}.

As the theoretical foundation of rank-$K$ regularization is not ready, it is unclear how to determine $\mathbf{B}$ and $\mathbf{\Xi}$ to minimize $\kappa(\mathbf{R})$.
Simply repeating the SR-$1$R would not work, as $\theta_{N-1}$ will keep decreasing.
After $K$ repetitions, $|\theta_{N-1}|$ will become extremely large and increase the condition number.
Therefore, in our example, we specifically look into three representative cases to reveal the potential of the rank-$K$ design.
\begin{itemize}
	\item \textbf{Approach~1}: 
	The first approach is to determine $\mathbf{B}$ and $\mathbf{\Xi}$ based on the EVD. 
	Recall \eqref{eq19} where $\mathbf{A}=\mathbf{U}\mathbf{\Lambda}\mathbf{U}^H$. 
	We move the first $K_1$ and last $K_2$ eigenvalues ($K_1+K_2=K$) to the average of the rest eigenvalues.
	In this case, $\mathbf{B}$ and $\mathbf{\Xi}$ are given by
	\begin{IEEEeqnarray}{rl}
		\mathbf{B}&=[\mathbf{u}_0,\cdots,\mathbf{u}_{K_1-1},\mathbf{u}_{N-K_2},\cdots \mathbf{u}_{N-1}],\\
		\mathbf{\Xi}&=\overline{\lambda}_{K_1,K_2}\mathbf{I}-\mathbf{\Lambda}_K,
	\end{IEEEeqnarray}
	where $\mathbf{u}_n$ stands for the $n^{th}$ column of $\mathbf{U}$, $\mathbf{\Lambda}_K$ for the diagonal matrix formed by the first $K_1$ and last $K_2$ eigenvalues, and $\overline{\lambda}_{K_1,K_2}$ for the average of other eigenvalues:
	\begin{equation}
		\overline{\lambda}_{K_1,K_2}=(\lambda_{K_1}+\cdots+\lambda_{N-K_2-1})/(N-K).
	\end{equation}

	\item \textbf{Approach~2}: 
	It is impractical to expect perfect EVD knowledge.
	Therefore, we add small Gaussian noise to $\mathbf{B}$ and $\mathbf{\Xi}$ to investigate the condition number when the knowledge of $\mathbf{B}$ and $\mathbf{\Xi}$ is not perfect. 
	\item \textbf{Performance Bound}: 
	The rank-$K$ regularization has been discussed in the literature \cite{Arbenz1988}, and we can conclude the following inequalities:
	\begin{IEEEeqnarray}{rl}
		\lambda_{n+K}\leq&\theta_{n}\leq\lambda_{n},_{~n=0,\cdots,N-K-1},\label{eq90}\\
		\lambda_{N-1}-\mathrm{Tr}(\mathbf{\Xi})\leq&\theta_{n}\leq\lambda_{n},_{~n=N-K,\cdots,N-1}.\label{eq91}
	\end{IEEEeqnarray}
	Similar to the discussions in Section~\ref{secIIIa}, \eqref{eq90} and \eqref{eq91} yield a lower bound of the condition number:
	\begin{equation}
		\kappa(\mathbf{R})\geq\lambda_{K}/\lambda_{N-K-1}.
	\end{equation}
	
\end{itemize}

\begin{figure}[t]
	\centering
	\includegraphics[scale=0.56]{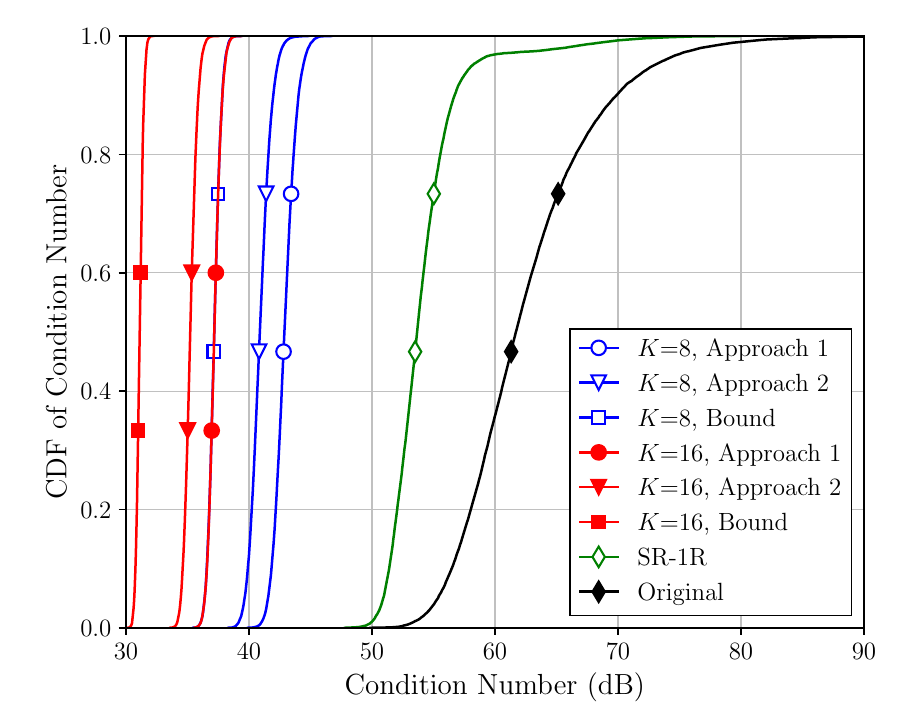}
	\caption{CDF of condition number for the rank-$K$ regularization in i.i.d. Rayleigh channel when $M=N=512$.}
		\vspace{-2em}
	\label{figRankK}
\end{figure}

In simulation, we consider $\mathbf{H}$ conforming to i.i.d. Rayleigh fading with $M=N=512$.
We consider $\mathbf{A}=\mathbf{H}^H\mathbf{H}$. 
For Approach~2, The Gaussian noise added to $\mathbf{B}$ and $\mathbf{\Xi}$ is i.i.d. with a variance of $0.015$.
We consider two cases: $K=16$ and $K=8$ where $K_1=K_2=(K)/(2)$.

Fig.~\ref{figRankK} shows the CDF of the condition number after rank-$K$ regularization.
As baselines, the green and black lines stand for the SR-$1$R and the original $\mathbf{A}$, respectively.
It is shown that Approach~1 reduces the condition number for around $12$~dB when $K=8$.
When $K$ increases to $16$, the condition number is reduced for around another $10$ dB.
Moreover, surprisingly, Approach~2 outperforms Approach~1 for around $3$~dB both when $K=8$ and $K=16$.
This is indeed encouraging and reveals a more optimal design of $\mathbf{B}$ and $\mathbf{\Xi}$ remains to be investigated in the future.

\end{document}